\documentclass[journal=jctcce,manuscript=article]{achemso}
\usepackage{amsmath}
\usepackage{amssymb}
\usepackage[version=4]{mhchem}
\usepackage{xcolor}
\usepackage{hyperref}
\usepackage{mathtools,xparse}
\usepackage{placeins}

\newcommand{\ket}[1]{ \left| #1 \right \rangle}

\title{Critical Limitations in Quantum-Selected Configuration Interaction Methods}

\author{Peter Reinholdt}
\affiliation{Department of Physics, Chemistry and Pharmacy, University of Southern Denmark, Campusvej~55, DK--5230 Odense M, Denmark.}
\email{reinholdt@sdu.dk}

\author{Karl Michael Ziems}
\affiliation{Department of Chemistry, Technical University of Denmark, Kemitorvet Building 207, DK-2800 Kongens Lyngby, Denmark.}
\alsoaffiliation{School of Chemistry, University of Southampton, Highfield, Southampton SO17 1BJ, United Kingdom.}

\author{Erik Rosendahl Kjellgren}
\affiliation{Department of Physics, Chemistry and Pharmacy, University of Southern Denmark, Campusvej~55, DK--5230 Odense M, Denmark.}

\author{Sonia Coriani}
\affiliation{Department of Chemistry, Technical University of Denmark, Kemitorvet Building 207, DK-2800 Kongens Lyngby, Denmark.}

\author{Stephan P. A. Sauer}
\affiliation{Department of Chemistry, University of Copenhagen, DK-2100 Copenhagen \O, Denmark.}
\author{Jacob Kongsted}

\affiliation{Department of Physics, Chemistry and Pharmacy, University of Southern Denmark, Campusvej~55, DK--5230 Odense M, Denmark.}
\begin{document}

\begin{tocentry}
\includegraphics[width=8.25cm]{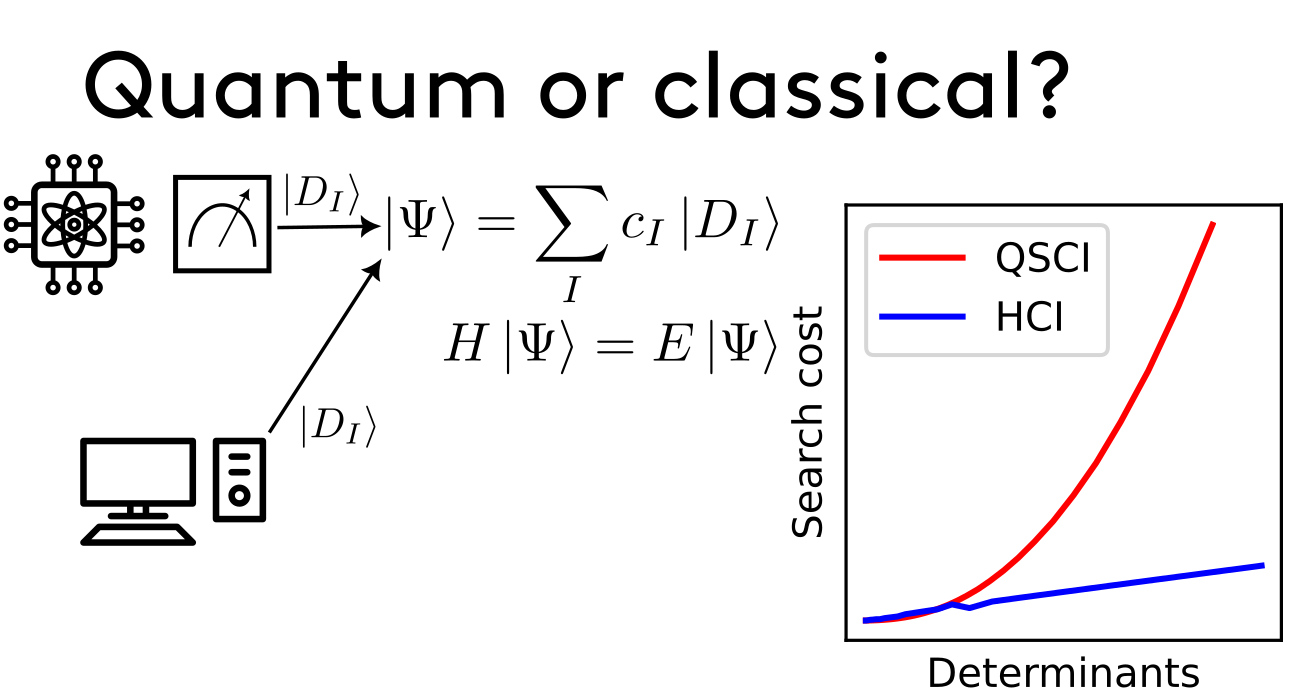}
\end{tocentry}

\begin{abstract}
Quantum Selected Configuration Interaction (QSCI) methods (also known as Sample-based Quantum Diagonalization, SQD) have emerged as promising near-term approaches to solving the electronic Schr{\"o}dinger equation with quantum computers. In this work, we perform numerical analysis to show that QSCI methods face critical limitations that severely hinder their practical applicability in chemistry. Using the nitrogen molecule and the iron-sulfur cluster [2Fe-2S] as examples, we demonstrate that while QSCI can, in principle, yield high-quality configuration interaction (CI) expansions similar to classical SCI heuristics in some cases, the method struggles with inefficiencies in finding new determinants as sampling repeatedly selects already seen configurations. 
This inefficiency becomes especially pronounced when targeting high-accuracy results or sampling from an approximate ansatz.
In cases where the sampling problem is not present, the resulting CI expansions are less compact than those generated from classical heuristics, rendering QSCI an overall more expensive method. 
Our findings suggest a significant drawback in QSCI methods when sampling from the ground-state distribution as the inescapable trade-off between finding sufficiently many determinants and generating compact, accurate CI expansions. This ultimately hinders utility in quantum chemistry applications, as QSCI falls behind more efficient classical counterparts. 
\end{abstract}

\maketitle

\section{Introduction}

The full configuration interaction (FCI) method provides the exact solution to the electronic Schr{\"o}dinger equation (within a given basis) and is thus considered the ground truth within quantum chemistry\cite{loos2020quest,eriksen2020ground,eriksen2020shape}. 
However, due to the exponential scaling in computing and memory requirements, exact FCI is limited to very small molecules. 
One of the largest demonstrations to date was by \citeauthor{gao2024distributed} on propane (C$_3$H$_8$)/STO-3G\cite{gao2024distributed}, who used a distributed FCI implementation on 256 nodes to treat a CI expansion of 1.3 trillion determinants.
Recent work using a small-tensor-product distributed active space representation has even reported calculations exceeding a quadrillion determinants \cite{shayit2025breakingquadrilliondeterminantbarrier}.
As a technical achievement, such work is certainly impressive.
If one is willing to settle for near-exact FCI results (within, e.g., micro-Hartree precision), approximate methods such as selected configuration interaction (SCI) can obtain results much more affordably\cite{loos2024go}.
Various developments within electronic structure theory have focused on achieving near-FCI accuracy at a lower cost\cite{eriksen2020shape}, including methods based on the density matrix renormalization group (DMRG)\cite{chan2011density}, many-body expanded FCI (MBE-FCI)\cite{eriksen2017virtual,eriksen2018many,eriksen2019many}, and various flavors of SCI\cite{huron1973iterative,tubman2016deterministic,holmes2016heat,liu2016ici}.
Recently, such methods were put to the test in a ``blind challenge'' of predicting the ground-state energy of the benzene molecule (C$_6$H$_6$) in a cc-pVDZ basis set\cite{eriksen2020ground}. It revealed that a firm consensus between various near-exact FCI methods has still not been reached, even when agreement from calculations on smaller systems would have suggested so.

Applying quantum computers within the field of quantum chemistry has received significant attention in recent years\cite{babbush2018low,mcardle2020quantum,lin2021real,klymko2022real}. This is partly due to the expectation that quantum computers should be a natural candidate for solving the electronic Schr{\"o}dinger equation as they inherently operate within an exponentially large Hilbert space.
In the long term, provided fault-tolerant devices become a reality, quantum computers could, for example, be used to obtain ground state energies using quantum phase estimation\cite{kitaev1995quantum,aspuru2005simulated}. 
However, current near-term quantum computers are noisy and rather limited in terms of qubit count, error rates, and execution speed, which limits their practical advantage over classical computing architectures\cite{zhou2020limits}. This has motivated the development of hybrid algorithms that take advantage of both quantum and classical computing architectures, resulting in shallower quantum circuits, which are more realistic options for practical quantum computing applications in the near term\cite{nash2020quantum,lau2022nisq,bauer2016hybrid}.
The variational quantum eigensolver (VQE) algorithm\cite{peruzzo2014variational,mcclean2016theory,tilly2022variational} has emerged as a particularly popular near-term method.
VQE-type methods rely on a parameterized ansatz, which is iteratively optimized using a hybrid quantum-classical feedback loop. The quantum device prepares a state and measures the expectation values over Pauli operators. The classical computer collects these measurements to evaluate expectation values over the fermionic Hamiltonian and updates the parameters of the parameterized ansatz using classical optimization methods.
Although popular, VQE-type methods have several serious drawbacks related to the measurement process,\cite{wecker2015progress} which inherently has statistical noise, and the expressibility of the ansatz\cite{liu2024mitigating,mcclean2018barren,tilly2022variational}.

% what is the state of the field with QSCI/SQD
The quantum selected configuration interaction (QSCI) method, introduced by \citeauthor{kanno2023quantum}\cite{kanno2023quantum,nakagawa2023adapt}, addresses some of the challenges faced with VQE by drastically reducing the quantum workload, leveraging quantum devices solely for determinant selection.
The central idea of the method is that some approximate ground state is prepared on a quantum computer, which is then used to draw samples corresponding to determinants in the computational basis. The CI Hamiltonian is then classically diagonalized in the basis of the selected determinants using conventional SCI diagonalization routines. We note that the classic diagonalization eventually becomes the computational bottleneck of the algorithm, provided enough determinants are discovered.
Thus, the ``quantum'' part of QSCI lies solely in the selection of determinants for the SCI expansion.
Such a division of work between quantum and classical resources has some clear advantages. First, since the quantum device is only used for selection, while energies are obtained classically, one removes the statistical errors associated with Hamiltonian measurement.
Second, the CI vector is accessible on the classical device after the diagonalization, which allows computing expectation values over other quantum mechanical operators or correcting the energy of the variational wave function with Epstein-Nesbet perturbation theory\cite{garniron2017hybrid}.
In the original QSCI work, the dimension of the subspace (i.e., the number of determinants included in the CI expansion) was rather limited, with at most around a hundred determinants, in part due to a limited variational space when using rather small basis sets.

The QSCI method has also been demonstrated on a larger scale by \citeauthor{robledo2024chemistry}\cite{robledo2024chemistry}, under the name sample-based quantum diagonalization (SQD) with applications to potential energy curves of the nitrogen molecule (in a cc-pVDZ basis) and the iron-sulfur cluster model systems [2Fe-2S] and [4Fe-4S], with active spaces of (30e,20o) and (54e,36o), respectively. 
The demonstrations in  Ref. \citenum{robledo2024chemistry} are relatively ``large-scale'' in nature, using CI expansion with millions of determinants.
The central procedure is similar to the original QSCI paper but demonstrates the use of a large number of qubits (up to 77) and handling aspects such as treating quantum samples with incorrect particle numbers.
The state preparation is handled with a local unitary cluster Jastrow ansatz (LUCJ)\cite{matsuzawa2020jastrow,motta2023bridging}, with parameters assigned classically based on CCSD $t_2$-amplitudes.
The SQD approach has since become a showcase for quantum-centric supercomputing\cite{alexeev2024quantum} and has been applied in several follow-up works, including the treatment of interaction energies with the potential energy surfaces of water and methane dimers\cite{kaliakin2024accurate}, spin state energetics of Fe(III)-NTA \cite{nutzel2024solving} and methylene\cite{liepuoniute2024quantum}, the treatment of excited states\cite{barison2025quantum}, inclusion of environment effects through density matrix embedding theory\cite{shajan2024towards}, in combination with implicit-solvation models\cite{kaliakin2025implicit}, and combined with phaseless auxiliary-
field quantum Monte Carlo\cite{yoshida2025auxiliary,danilov2025enhancing}.

Separately, we note that an approach for determining the ground state of the SCI using quantum computers has also been proposed\cite{yoffe2025qubit}. Unlike QSCI and SQD, determinants are selected using conventional SCI heuristics, and the focus is on leveraging variational quantum circuits to find the ground-state energy of the CI Hamiltonian.

The QSCI method introduces a potentially useful approach to determinant selection via quantum sampling, but several fundamental challenges limit its practical utility in real-world applications.  
First of all, the method arguably requires the preparation of a good state on the quantum device. Ideally, such a state should provide a probability distribution that closely matches the exact FCI solution since one would otherwise miss important determinants or include a significant fraction of irrelevant ones. Preparing such states is difficult, and it remains unclear how such a state preparation could take place without solving the electronic Schr{\"o}dinger equation to high accuracy in the first place.
Independently, two groups recently proposed methods with state-preparation protocols based on time-evolved or Hamiltonian-simulation QSCI, which could perhaps provide a workaround for this problem\cite{sugisaki2024hamiltonian,mikkelsen2024quantum}. Later, similar developments under the SQD moniker have independently appeared\cite{yu2025quantumcentricalgorithmsamplebasedkrylov}.

A second issue arises from the noisy nature of currently available quantum devices. Even though the target ansatz is, in principle, particle-conserving (i.e., has a desired number of $\alpha$ and  $\beta$ electrons), the presence of hardware noise leads to many samples having the wrong number of particles. For example, in 
Ref.~\citenum{robledo2024chemistry}, only 0.44\% of samples for the iron-sulfur cluster [2Fe-2S] (30e,20o) are within the target particle sector. Solutions to this problem have been proposed in the form of methods like the \emph{self-consistent configuration recovery} (S-CORE)\cite{robledo2024chemistry}, which seek to recover nearby bit-strings with a correct particle number from the set of erroneous bitstrings.

In our work, we will assume that these problems could be solved by a combination of improvements in quantum hardware and algorithms. Thus, we will not consider the state preparation or the noise problem but rather assume that an exact ansatz can be found. To this end, we rely on determinants sampled from CASCI wave functions. This approach eliminates complications related to noisy quantum devices, allowing us to focus on assessing the intrinsic efficiency and the quality of QSCI expansions.
We focus on sampling from a ground-state distribution as this is the distribution that has mainly been targeted in previous SQD/QSCI works and, crucially, for this distribution, practical approximate state preparation strategies are known, such as VQE or Krylov-based approaches. As we shall discuss later, it might be theoretically possible to construct states that yield more efficient sampling. Yet, as of now, this is mostly hypothetical, as to the best of our knowledge, no method exists for efficiently preparing such states.

Ultimately, we demonstrate that even with an ideal sampling wave function in a noise-free environment, the stochastic nature of sampling itself either introduces significant inefficiencies or leads to less compact wave functions. Less compact CI expansions (i.e., more determinants for the same energy) lead to higher (classical) computational costs associated with the diagonalization of the CI Hamiltonian for the same final accuracy in the computation. These limitations undermine any practical advantage of QSCI over well-established classical SCI heuristics. Furthermore, enhancements like ext-SQD\cite{barison2025quantum}, which add determinants post-sampling, could equally be applied to classical SCI methods, negating any unique benefit of quantum sampling.

\section{Theory}

A configuration interaction (CI) wave function can be expressed as a linear combination of Slater determinants, $\ket{\Phi_I}$,
\begin{equation}
    \ket{\Psi} = \sum_{I} c_I \ket{\Phi_I},
\end{equation}
with the CI coefficients, $c_I$. The included determinants can be selected by excitation level out of a reference determinant (e.g., Hartree-Fock), forming the CISD, CISDT, etc. hierarchy\cite{siegbahn1992configuration}, which eventually terminates at the FCI expansion when $N$-tuple excitations are included.
Alternatively, as is done in SCI expansions\cite{huron1973iterative,tubman2016deterministic,holmes2016heat,liu2016ici}, a smaller set of \emph{important} determinants can be selected based on classical heuristics. This approach recognizes that only a small fraction of determinants in the FCI expansion contribute significantly to the electronic wave function for most molecular systems \cite{ivanic2001identification}.

The electronic Schr{\"o}dinger equation is solved in the subspace spanned by the chosen Slater determinants via diagonalization of the CI Hamiltonian matrix 
\begin{equation}
H_{IJ} = \left\langle \Phi_I \left| \hat{H} \right| \Phi_J \right\rangle,
\end{equation}
where $H_{IJ}$ are matrix elements of the electronic Hamiltonian $\hat{H}$. % between determinants $\ket{\Phi_I}$ and $\ket{\Phi_J}$. 
The eigenvalues of the CI Hamiltonian matrix correspond to the electronic energy levels, and the eigenvectors (or CI vectors) contain the CI coefficients for the associated wave functions.  
The CI coefficients have a direct connection to the probability amplitudes for some particular determinant $\ket{\Phi_I}$, with the probability given by the norm-square of the CI coefficient, $|c_I|^2$. This probabilistic interpretation forms the basis of quantum sampling techniques, such as those used in QSCI, where determinants are stochastically selected by drawing samples from a wave function prepared on the quantum device.

In addition to QSCI, we will also consider a representative of classical SCI. In particular, we will consider Heat-bath Configuration Interaction\cite{holmes2016heat} (HCI). In HCI, important determinants for the CI expansion are added based on their expected contribution to the wave function. It uses a parameter $\varepsilon$ to screen a potential determinant $\ket{\Phi_I}$ and includes it if any of the existing determinants in the SCI expansion satisfy
\begin{equation}
\left|H_{IJ} c_J \right| > \varepsilon. \label{eq:hci_criterion}
\end{equation}
Here, $H_{IJ}$ is the Hamiltonian matrix element between an already included determinant $\ket{\Phi_J}$ and the candidate determinant $\ket{\Phi_I}$, while $c_J$ is the coefficient of the included determinant $\ket{\Phi_J}$. This ensures efficient selection by focusing on determinants with significant coupling, typically yielding compact and accurate CI expansions.

The HCI method works in an iterative fashion. Starting from some initial CI expansion (e.g., the HF determinant), the following steps are repeated:
1) the CI Hamiltonian is diagonalized, and the CI coefficients of the current determinant set are obtained;
2) determinants are added according to the criterion in eq. \eqref{eq:hci_criterion}, with 
$\varepsilon$ as a parameter.
These steps are repeated until no more determinants can be added according to eq. \eqref{eq:hci_criterion} (in the original work\cite{holmes2016heat}, when the number of new determinants is less than 1\% of the number of determinants already selected).

\section{Computational Details}\label{sec:computational_details}
Hartree-Fock and CASCI (10e, 22o) wave functions, as well as one- and two-electron molecular orbital (MO) integrals (cc-pVDZ\cite{dunning1989gaussian} basis) of the nitrogen molecule, were obtained using PySCF \cite{pyscf}. 
We considered structures near the equilibrium geometry ($R=1.09$\AA) and along the potential energy surface $R=0.7$\AA~to $R=3.0$\AA.
Additionally, we investigated the iron-sulfur cluster [2Fe-2S] (30e,20o), where the one- and two-electron integrals were taken from the FCIDUMP file published by \citeauthor{li2017spin}\cite{li2017spin,Li2021zhendongli2008}.
Finally, we also include results to assess the scalability of QSCI with system size, using linear lithium hydride chains, $(\mathrm{LiH})_n$/STO-3G, as an example, applying full-valence CAS($2n$e, $5n$o) active spaces. For this, we consider a weakly correlated regime $R_{\mathrm{Li-H}}=1.61$ \AA, $R_{\mathrm{LiH-LiH}}=1.81$ \AA, and a strongly correlated regime with $R_{\mathrm{Li-H}}=3.22$ \AA, $R_{\mathrm{LiH-LiH}}=3.62$ \AA.
Throughout the paper, we adopted the notation ($N$e, $M$o) for the active spaces, where $N$ is the number of active electrons and $M$ is the number of active orbitals.

The QSCI calculations were performed by drawing determinant samples from the CI vectors from the CASCI calculations with the probability of selecting a particular determinant given as $|c_I|^2$. %This represents how an ansatz is measured probabilistically on a quantum device in the computational basis. 
The number of samples corresponds to the number of times the ansatz is measured. Our approach constitutes an ideal limit of QSCI since it assumes perfect state preparation of the exact CI vector and avoids any quantum device noise (i.e., non-ideal execution of the quantum circuits).
The PyCI library\cite{richer2024pyci} was used to diagonalize the resulting CI Hamiltonian. 
HCI calculations were also performed using PyCI.

To assess the impact of an approximate ansatz in the sampling routine of QSCI, we obtained state vectors of equilibrium N$_2$ for the unitary cluster Jastrow (UCJ)\cite{matsuzawa2020jastrow} and local UCJ (LUCJ)\cite{motta2023bridging} ansatz using ffsim\cite{ffsim}. For both methods, two layers were used, and the ansatz parameters were assigned using the $t_2$ amplitudes of a classic CCSD simulation. This mimics the state preparation procedure used in recent works of IBM Quantum employing SQD/QSCI on quantum hardware\cite{robledo2024chemistry}. 

\section{Results and Discussion}
We evaluate the performance of the QSCI method on two distinct systems: the N$_2$ molecule and the [2Fe-2S] iron-sulfur cluster. The N$_2$ molecule is a prototypical example of a system dominated by weak, dynamic correlation, with static correlation effects eventually becoming significant as the triple bond is stretched. In contrast, the [2Fe-2S] cluster represents a more strongly correlated system, providing a challenging test for the method in capturing multi-reference effects. By comparing QSCI to classical heuristics like HCI for these cases, we can assess the efficiency and accuracy of the CI expansions produced by the QSCI method.

\subsection{Potential energy curves of the nitrogen molecule}

We first consider the ground state potential energy curve of N$_2$/cc-pVDZ (10e, 22o). This active space is sufficiently small to allow CASCI reference calculations, which involve $0.693 \times 10^9$ determinants in the CASCI expansion.
We use the CI vector from the CASCI wave function to provide the QSCI samples.
With this definition, QSCI will recover the full correlation energy from the reference CASCI wave function in the limit of infinite sampling. 

\begin{figure}
    \centering
    \includegraphics[width=1.0\textwidth]{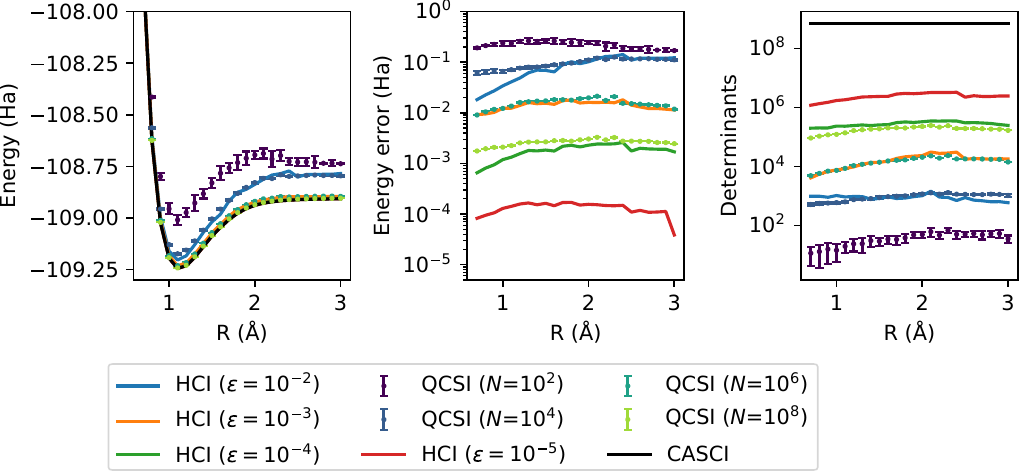}
    \caption{Ground-state potential energy curve of N$_2$/cc-pVDZ (10e,22o). The left panel shows total energies from HCI (solid lines), QSCI (dots), and CASCI (solid black line) wave functions with different thresholds $\varepsilon$ (HCI) and number of samples $N$  (QSCI). The middle panel shows energy errors relative to the CASCI reference. The right panel shows the number of determinants in the CI expansions. For the QSCI results, we include ten independent runs and indicate the variance across runs with error bars at a 95\% confidence interval.}
    \label{fig:n2_pes}
\end{figure}

As shown in Figure \ref{fig:n2_pes}, we find a systematic convergence towards the reference CASCI energy when tightening the threshold $\varepsilon$ for the HCI wave functions. At $\varepsilon=10^{-3}$, the potential energy curve is in qualitative agreement with the reference CASCI result, and energy errors on the order of one milli-Hartree are reached with $\varepsilon=10^{-4}$.
Similarly, the QSCI wavefunction converges systematically towards the reference CASCI result with an increasing number of samples.
The QSCI wave function provides energy errors on the order of $10^{-2}$ Hartree with $10^6$ samples, which is quite similar to the HCI ($\varepsilon=10^{-3}$) result.
Errors on the order of one milli-Hartree can be reached after using about $10^{8}$ -- $10^{9}$ samples.
Due to the probabilistic nature of the QSCI sampling, different runs will generally select different sets of determinants. 
Thus, when a small number of samples is used, there is a noticeable variance in the energies from different runs. This highlights that, although the process of solving the SCI Hamiltonian with some particular set of determinants leads to an energy estimation with no inherent statistical variance (unlike, e.g., VQE), the selection of the determinants themselves in the QSCI process \emph{is} probabilistic, meaning that there is still statistical noise in the QSCI process. Fortunately, this variance becomes less noticeable when the number of samples is increased.

The rightmost panel of Figure \ref{fig:n2_pes} shows the number of determinants included in the CI expansion across the nitrogen potential energy surface. For both HCI and QSCI, the size of the CI expansion increases rapidly when tightening the threshold $\varepsilon$ (for HCI) or increasing the number of samples $N$ (for QSCI). Notably, we find that the HCI ($\varepsilon=10^{-3}$) result and QSCI with $10^6$ samples align quite well in terms of the number of determinants (and as mentioned before, in the energy). This suggests that, at least for this system, the quality of the determinants selected by QSCI is comparable to that obtained from classical heuristics (i.e., HCI), which is certainly encouraging.
However, a sampling problem is also apparent. With $10^6$ samples, we find around $15 \times 10^3$ determinants with QSCI. Increasing the number of samples to $10^8$, the number of determinants increases only to about $182\times 10^3$ determinants, which means that a 100-fold increase in the sampling yields only a 12-fold increase in the number of found determinants. 
This happens because the same determinants are repeatedly sampled, providing no new information for the QSCI process and leading to a high number of ``wasted'' quantum samples, translating directly to immense measurement overheads.
Coupled with the fact that obtaining good accuracy in the energy requires a significant number of determinants in the CI expansion, converging a QSCI wave function to high accuracy will be challenging.

\begin{figure}
    \centering
    \includegraphics[width=1.0\textwidth]{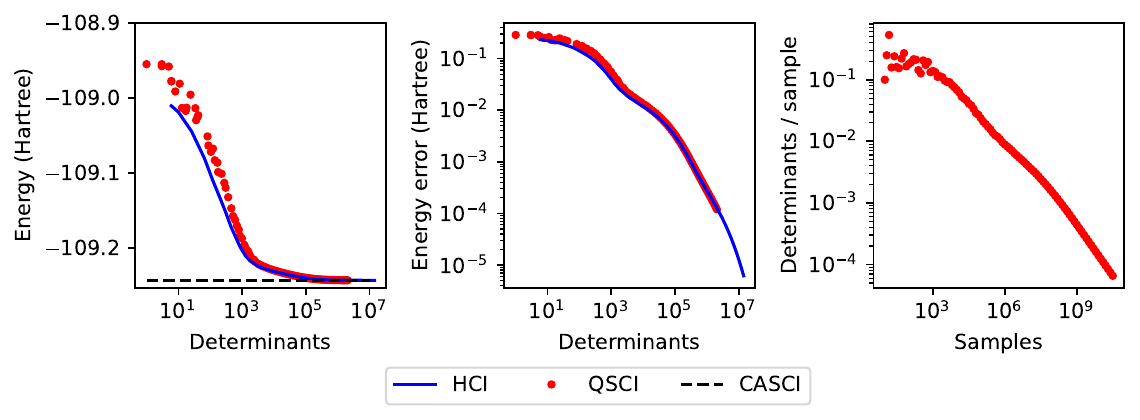}
    \caption{Ground-state energy of the N$_2$/cc-pVDZ molecule ($R=1.09$ \AA) with a (10e,22o) active space. The left panel shows total energies from HCI, QSCI, and CASCI wave functions as a function of the number of determinants included in the CI expansion. The middle panel shows energy errors relative to the CASCI reference. The right panel shows the determinants per sample for QSCI as a function of the number of samples.}
    \label{fig:n2_detail_energy}
\end{figure}

Next, we examine the sampling problem in more detail, focusing on N$_2$ at $R=1.09$ \AA.   
In Figure \ref{fig:n2_detail_energy}, we show the energy and error relative to the CASCI energy for wave functions produced by QSCI and HCI, plotting the resulting energy against the number of determinants of the respective CI expansions.
With HCI, each data point in the plot corresponds to a unique converged HCI calculation with increasingly tight $\varepsilon$, logarithmically spaced ($\varepsilon=10^{-1.0}, 10^{-1.1}, 10^{-1.2}, ..., 10^{-6.0}$).
Similarly, with QSCI, each point corresponds to a unique QSCI instance with an increasing number of samples, also logarithmically spaced ($10^{1.0}, 10^{1.1}, ..., 10^{10.5}$).
Clearly, the quality of HCI (blue lines) and QSCI (red dots) CI expansions are almost identical across a wide number of determinants, perhaps with a slim advantage to HCI.

However, as shown in the rightmost panel of Figure \ref{fig:n2_detail_energy}, the sampling issue of QSCI eventually becomes severe, with very few new determinants being discovered. We quantify this in terms of the number of determinants per sample (DPS), defined as $N_\text{determinants} / N_\text{samples}$.
Initially (at $10^3$ samples), sampling from the CI vector yields about 0.1 determinants per sample. As more determinants are found, already-found determinants are frequently sampled again. As a result, the DPS drops to 0.01 at $10^6$ samples and 0.0005 at $10^9$ samples. 
As shown in Figure S2, if one wanted to target micro-Hartree precision on N$_2$, it would require an estimated $4\times 10^{14}$ samples.

Another way to see the severity of the repeated sampling problem is to compare the search costs of QSCI and HCI when increasing the size (i.e., accuracy) of the CI expansion.
As shown in Figure S4, the problem of repeatedly finding the same determinants with QSCI manifests as a super-linear (nearly quadratic) search cost scaling for N$_2$. In contrast, HCI shows only a linear increase in the search cost as a function of the number of determinants in the CI expansion.
The HCI method progressively grows the CI expansion through successive iterations. In addition to determinant selection, this requires repeated diagonalizations of the SCI Hamiltonian. Thus, with HCI, one should also factor in the cumulative costs of diagonalization. In contrast, QSCI in an ideal setting requires only a single round of diagonalization.
Even when considering the total calculation costs of HCI, including determinant search, sparse Hamiltonian construction, and Davidson diagonalization, the observed scaling in determinants of HCI remains more modest at around $N_{\mathrm{det}}^{1.2}-N_{\mathrm{det}}^{1.3}$ for the total calculation time than just the search time of QSCI (excluding costs associated with diagonalization) at around $N_{\mathrm{det}}^{2.2}-N_{\mathrm{det}}^{2.9}$ (see Figures S4 and S5 and the surrounding discussion).
A direct comparison of computational timings requires assigning a conversion factor from quantum samples to seconds, which we will consider in Section \ref{sec:generality}.

\FloatBarrier

\subsection{Scaled probability distributions}
To understand this problem further, we consider results obtained by sampling from the scaled probability distribution
\begin{equation}
    \tilde{p}_I = \frac{(p_I)^{\alpha}}{\left(\sum_J (p_J)^\alpha\right)}
    \label{eq:pdf_scaling}
\end{equation}
where $\alpha$ is a scaling parameter and $p_I = |c_I|^2$ is the original CI vector probability of a given determinant $\ket{\Phi_I}$. This can be seen as maintaining the support (configuration space) of the FCI wave function but changing its weight distribution. Such a probability distribution allows us to tune the sampling by changing the weights of the determinants in the original CI vector, making it more likely (or less likely) to find new determinants depending on the value of $\alpha$.
The original distribution of the determinants is recovered with $\alpha = 1$. Letting $\alpha>1$ emphasizes the larger weights in the probability distribution, while letting $\alpha<1$ emphasizes smaller weights in the probability distribution. At $\alpha=0$, this distribution corresponds to uniform sampling from the correct particle sector (see also Figure S1).
We do not suggest such scaled distributions as practically implementable, but rather to serve as a stand-in for how probability distributions of more diverse molecular systems might look more generally.

\begin{figure}
    \centering
    \includegraphics[width=1.0\textwidth]{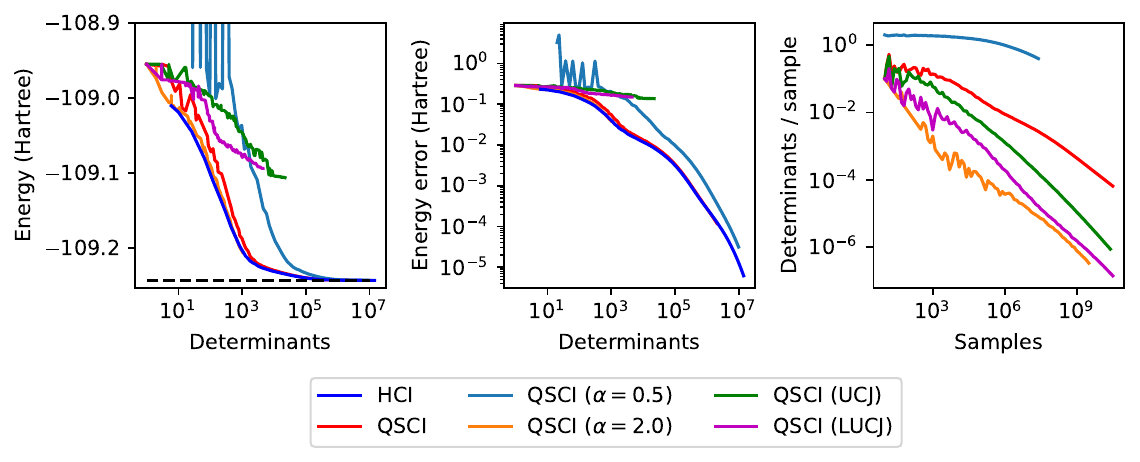}
    \caption{Ground-state energy of the N$_2$/cc-pVDZ molecule ($R=1.09$ \AA) with a (10e,22o) active space. The left panel shows total energies as a function of the number of determinants included in the CI expansion, with the black dashed lines indicating the CASCI energy.  
    The middle panel shows energy errors from HCI and QSCI wave functions (relative to the CASCI reference) as a function of the number of determinants included in the CI expansion. The right panel shows the determinants per sample for QSCI as a function of the number of samples. For QSCI, we include sampling from the original CASCI distribution, as well as scaled CASCI distributions (see Eq.~\eqref{eq:pdf_scaling}), and from UCJ and LUCJ wave functions.}
    \label{fig:n2_detail_extras}
\end{figure}

Such distributions are probably not realizable on real quantum hardware (except $\alpha=0$) but allow us to reason about the properties of alternative probability distributions while keeping the ranking of the CI weights intact (i.e., if $p_I>p_J$, the scaled distributions also satisfy $\tilde{p}_I>\tilde{p}_J$).
In Figure \ref{fig:n2_detail_extras}, we show the effect of the scaled distribution for our N$_2$ example. Letting $0 \leq \alpha<1$ will increase the frequency of finding determinants with small weights, alleviating the sampling problem but making the resulting CI expansions less compact.
Clearly, for $\alpha=0.5$ (light blue dots) in Figure \ref{fig:n2_detail_extras}, the sampling problem is essentially not present, with the DPS remaining high, even when going towards a large number of samples.
The resulting CI expansion is, for a low number of determinants, not very good (note the high-error cyan points in the left panel of Figure \ref{fig:n2_detail_extras}). As the number of samples increases, most of the important determinants become selected, and a reasonably good energy is obtained, albeit with much less compact CI expansion, with around twice the number of determinants for the same energy accuracy as HCI or QSCI (with an unmodified probability distribution, $\alpha=1$). It should be noted that the (reduced) compactness of the wave function translates directly to the costs of the classical diagonalization step, with too-large expansions eventually becoming computationally intractable. In fact, we need so many determinants for $\alpha=0.5$  that the memory and compute requirements in the solution of the SCI Hamiltonian begin to become significant.
As discussed above, whereas QSCI requires only a single diagonalization step (assuming ideal execution such that configuration recovery is not required), HCI, as an iterative process, requires repeated diagonalization and search steps, which adds a computational overhead to HCI compared to QSCI for a given compactness. 
In practice, one would likely still advise an approach for QSCI in which increasingly large diagonalisations are performed until some degree of convergence is obtained. 
The cost of diagonalization depends on the size of the CI dimension, which is small in the initial HCI iterations, making the early diagonalizations computationally inexpensive.
In Table S1, we provide a concrete example of the total diagonalization costs for HCI, QSCI, and the scaled QSCI distribution with alleviated sampling efficiency ($\alpha=0.5$) for a fixed target energy accuracy. 
Our results show that the total computational cost of HCI is comparable to the diagonalization cost alone of the scaled QSCI distribution ($\alpha=0.5$), indicating that the overhead from repeated diagonalization in HCI is manageable.
We also estimated the computational cost related to quantum sampling by including a conversion factor from quantum samples to seconds in the timings (i.e., a quantum sampling rate).
Even with relatively optimistic estimates for the sampling rate of $10^6$ quantum samples per second, we find that the computational cost of HCI is lower than both the scaled ($\alpha=0.5$) and unscaled versions of QSCI. 

Thus, while alternative input distributions might alleviate the sampling problem, the overall efficiency of the QSCI method would not necessarily improve if they result in less compact CI expansions.

When the frequency of finding determinants with large weights is enhanced ($\alpha>1$), finding new determinants is difficult, but any selected determinants will most likely contribute significantly to the CI expansion.
Thus, for $\alpha = 2.0$ (see Figure \ref{fig:n2_detail_extras}, orange dots), we observe more severe sampling issues manifesting as very low DPS already at a low number of samples. However, the resulting CI expansions are very compact, even outperforming HCI by a slim margin. For example, using QSCI ($\alpha=2.0$), we obtain an energy error of 0.037 Hartree with 1050 determinants (from $2.5 \times 10^9$ samples), where similar accuracy with HCI required 1100 determinants. Standard QSCI ($\alpha=1.0$) requires more determinants (around 1500), i.e., it is less compact but finds these determinants with only about 40,000 samples for a similar error.
These results using artificially weighted distributions highlight an essential and fundamental issue with selecting determinants by sampling: the ground-state probability distribution can either be good at discovering many new determinants or good at selecting a compact CI expansion, but not both at the same time.
Further scaling factors are shown in Figure S3.

An additional factor to consider is that preparing an FCI-quality wave function on the quantum computer is difficult in the first place.
In practice, the determinants would likely be sampled from a lower-quality wave function.
We include examples of such wave functions in Figure \ref{fig:n2_detail_extras}, exemplified by the 2-UCJ\cite{matsuzawa2020jastrow} and heavy-hex 2-LUCJ\cite{motta2023bridging} wave functions, with initial parameters assigned from CCSD $t_2$-amplitudes. This choice of ansatz has proven popular for previous hardware demonstrations\cite{robledo2024chemistry}, although we point out that not a full 2-layer LUCJ was used in these works but rather a more approximated approach, resembling an ansatz structure between one and two layers.
We find that sampling from the UCJ and LUCJ wave functions gives significantly worse results than the previously discussed sampling from the CASCI wave function. This manifests both in terms of a lower number of determinants per sample and significantly worse energies, even for relatively large CI expansions. The worse energies are likely due to a different ranking of the determinant weights, as shown in Figure S6. Curiously, LUCJ has a slightly better determinant selection than UCJ, although this is coupled with a more severe difficulty in finding new determinants and has an overall worse energy than QSCI sampled from the ideal CI vector.
The less-than-stellar performance of the approximate wave functions (UCJ/LUCJ) could, at least in principle, be improved by actually optimizing the ansatz rather than just assigning parameters based on CCSD $t_2$ amplitudes. Such optimization has been investigated to some extent in Ref. \citenum{robledo2024chemistry} using quantum simulators. However, this optimization is likely to bring the probability distribution closer to the ground-state distribution, which, as shown above, can still lead to inefficient sampling. To the best of our knowledge, implementing this optimization on quantum hardware in practice proves to be a significant challenge and would still suffer from the fundamental problems of QSCI reported above.
In general, the structure and parameter initialization of UCJ/LUCJ ans\"atze are not yet well-understood. For example, in Ref. \citenum{danilov2025enhancing}, the authors find that LUCJ circuits with completely \emph{random} parameter initializations performed better than the more established approach of assigning parameters based on CCSD $t_2$-amplitudes. 
Thus, the poorer performance of LUCJ compared to exact ground-state sampling in the present study should not be taken as a definitive indication of its potential, and the effectiveness of such approximate ans\"atze could improve as theoretical understanding advances.

\subsection{More correlated systems: Iron-sulfur clusters}

Next, we consider the more challenging electronic structure of the iron-sulfur cluster [2Fe-2S] in a (30e,20o) active space. The one- and two-electron integrals for this system were obtained from \citeauthor{li2017spin}\cite{li2017spin,Li2021zhendongli2008}.
We note that the same system has been studied in several QSCI works\cite{robledo2024chemistry,barison2025quantum}. 
The number of determinants in this space ($240\times 10^6$) is small enough that a CASCI wave function can comfortably be obtained, so we will use this as the sampling wave function for the QSCI.
\begin{figure}
    \centering
    \includegraphics[width=1.0\textwidth]{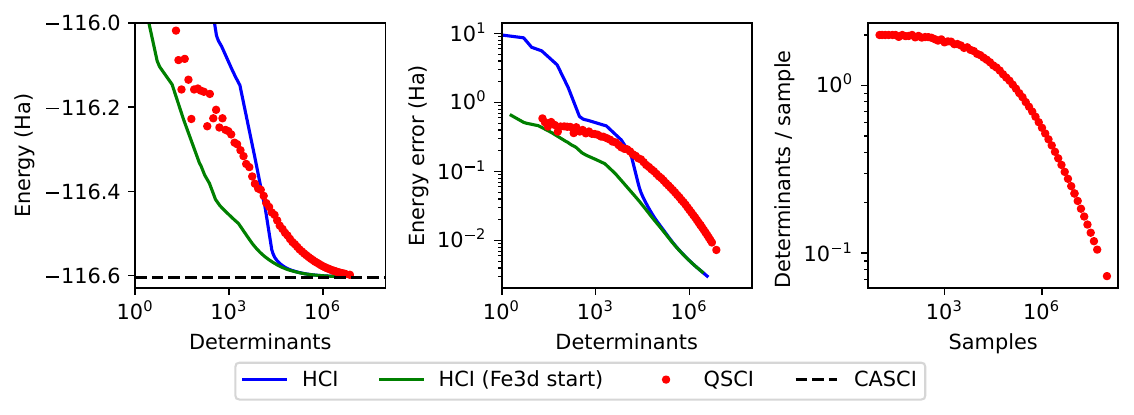}
    \caption{Convergence in the total energy for the HCI and QSCI approaches for the [2Fe-2S] system. The left panel shows the total energy as a function of selected determinants. The CASCI energy is given as a reference (dashed line). The CASCI wave function contains $240\times 10^6$ determinants. The middle panel plots the error in the energy relative to the CASCI reference. The right panel displays the number of determinants per sample for QSCI. For HCI, we include results starting from an HF-like determinant (blue lines, ``bad guess'') or determinants with anti-ferromagnetically coupled spins on the iron centers (green lines, ``good guess'').}
    \label{fig:fe2s2_energy}
\end{figure}
Figure \ref{fig:fe2s2_energy} shows the convergence in the energy as a function of the number of determinants included in the CI expansions for HCI and QSCI wave functions. For the HCI results, we include results for two different initial guesses. First, starting from an HF-like determinant (blue lines), which constitutes the simple yet ``bad guess''. As the orbitals from Ref. \citenum{li2017spin} are local orbitals, the HF-like determinant corresponds to filling all electrons into the left side of the iron-sulfur cluster, which is clearly an unphysical configuration. In our second approach, we start from a set of $M_s=0$ determinants (green lines) with five spin-up electrons on one of the iron centers (Fe 3d orbitals) and five spin-down electrons on the other iron center. This corresponds to a configuration of anti-ferromagnetically coupled spins and would be a ``good guess'' based on a chemical understanding of the system. It also happens to be the largest weight in the CASCI expansion.
The initial guess also impacts the magnitude of the ``correlation energy'', defined as the energy difference between the single-determinant Hartree-Fock and the CASCI result. For the ``bad guess'', one obtains a correlation energy of $-9.5$ Hartree, while for the chemically intuitive anti-ferromagnetically coupled set of determinants, one finds a smaller correlation energy of just $-0.65$ Hartree.
In Figure \ref{fig:fe2s2_energy}, we show the HCI results and see that the HCI expansions are also sensitive to the choice of starting determinants, with chemical intuition allowing for more compact CI expansions. For those of us who lack good chemical intuition, the good news is that after including a sufficiently large number of determinants (after around $10^5$), the two HCI calculations attain comparable energies, even when a bad initial guess is used.
With the bad HCI starting guess (blue lines), the QSCI wave function (red lines) is initially (i.e., at large energy errors) more compact than the HCI one until a cross-over point at around 10,000 determinants (at an energy error of about 0.2 Ha). Beyond this point, for the same accuracy result, the QSCI wave function always requires more determinants than the HCI expansions.
The difference is even more stark when comparing QSCI with the good starting guess of HCI expansion (green lines). Here, the QSCI wave function is consistently less compact. For example, HCI manages an energy error below 0.1 Hartree with 4,260 determinants, which is more than one order of magnitude smaller than the 50,551 determinants required by QSCI for the same energy.

On the other hand, the problem of repeated sampling of already-seen determinants is less severe for the [2Fe-2S] system. Initially, almost every quantum sample leads to a new determinant (two, if the $\alpha$ and $\beta$-strings of a sample are not identical). As sampling progresses, the number of new determinants per sample begins to level off.  As a result, at $10^8$ samples, only around $10^7$ determinants are found. The problem of repeated sampling is much less pronounced than that of the nitrogen molecule since the CASCI wave function amplitudes have a rather flat and slowly decaying distribution with no single large weight (compare blue and black lines in Figure S7). This makes finding unseen determinants relatively easier. However, finding many determinants is not the primary goal of a selection algorithm for SCI.  Rather, it is finding \emph{good} determinants, which allow for a compact yet accurate CI expansion. As evidenced by the results in Figure \ref{fig:fe2s2_energy}, QSCI is not optimal in this respect and will thus need significantly more computational resources for the classical diagonalization step. 

A careful reader may have noticed that our CI expansions (both HCI and QSCI) are much smaller than those reported in previous QSCI works (see, for example, Ref. \citenum{robledo2024chemistry} and Figure S10 therein). In Ref. \citenum{robledo2024chemistry}, on the exact same [2Fe-2S] system, around $100\times10^6$ determinants are required to reach an energy of around $-116.60$ Hartree, more than two orders of magnitude greater than the around $0.81\times 10^6$ determinants required in our calculations. The reportedly large number of determinants was especially surprising to us, considering that the entire CASCI space contains only $240\times 10^6$ determinants. This means that in some of the larger calculations of Ref. \citenum{robledo2024chemistry}, which include nearly $200\times10^6$ determinants, more than 80\% of all determinants have been selected for the SCI expansion.
Although the sampling component is not entirely identical, the (classically) variationally optimized LUCJ circuits employed in Ref. \citenum{robledo2024chemistry} likely approximate the ground-state distribution we consider reasonably well. 
Thus, we believe the discrepancy could be related to a slightly suboptimal SCI solver. A cursory glance at the implementation\cite{Johnson2024Qiskit} reveals that the approach relies on the \texttt{kernel\_fixed\_space} selected CI subroutine from PySCF\cite{pyscf}. As mentioned in the comments in the source code of the PySCF implementation, this subroutine is ``\emph{...an inefficient dialect of Selected CI using the same structure as determinant based FCI algorithm}''. In particular, this implementation does not select individual determinants. Instead, a set of $\alpha$- and $\beta$-strings are selected, and every determinant formed by all combinations thereof is included in the CI expansion. Such an algorithm may include up to quadratically more determinants in the CI expansion compared to individually selecting determinants. While this partially alleviates the sampling problem (since many more determinants are included), it somewhat contradicts the core aim of selected CI expansions -- namely, to obtain good wave functions with as few determinants as possible. 
Notably, our results challenge the claim that for [2Fe-2S], quantum sampling is more efficient than the classic HCI method\cite{robledo2024chemistry}.

\subsection{Generality and Sampling Considerations}\label{sec:generality}
%Next, we address the question of generality and modifications to the originally proposed sampling support. 
%
So far, we have considered two specific systems and a class of scaled distributions for sampling. With these systems, we found a clear trade-off between sampling efficiency and the compactness of the resulting CI expansion. Strengthening the generality of our results, in the Supporting Information, we provide further examples of 29 additional chemical systems derived from the W4-17-MR benchmark set\cite{karton2017w4} with some additional hand-picked systems (see Figures S8-S36).
The conclusions drawn so far appear to be valid across a range of chemical systems, with the typical obstacle being poor sampling efficiency (similar to N$_2$).
Of interest are systems such as B$_2$ and twisted ethene. These systems exhibit a behavior similar to the [2Fe-2S] system in the sense that there is an intermediate region where QSCI can provide more compact CI expansions than HCI initialized from the HF determinant, although this advantage is eventually lost when reaching more sizeable CI expansions, i.e., more accurate energies. A common feature of these systems is that the Hartree-Fock determinant is a poor initial guess for HCI as it has negligible weight in the ground-state CI expansion. Therefore, when HCI is initialized from the HF determinant, there is an intermediate region with the possibility of converging to excited states. However, as the threshold $\varepsilon$ is lowered, eventually, one connects to determinants important for the ground state, which leads to rapid convergence to the ground-state energy.
These drawbacks can sometimes be avoided with good chemical intuition (recall [2Fe-2S]) or starting the HCI from determinants selected from a CIS calculation (twisted ethene).

In our work, we have generally assumed that the sampling from QSCI should occur from the ground-state distribution, in line with the original proposal of QSCI\cite{kanno2023quantum,nakagawa2023adapt}. 
Arguably, this is also the implicit assumption in the larger-scale SQD studies\cite{robledo2024chemistry}, which have employed optimization-free LUCJ with parameter initialization from CCSD $t_2$-amplitudes for their quantum experiments. However, the authors note that ``\emph{quantum-classical optimization could further improve the quality of the solutions.}'' and present in the supporting information simulator studies 
with LUCJ circuits ``\emph{optimized to minimize the estimator energy}'' and perform in-depth analysis on the \emph{ground state} wave function concentration for quantum sampling. Optimization (i.e., energy minimization) of such ans\"atze would, in all likelihood, bring the probability distributions closer to the FCI distribution.
Our work has demonstrated clear sampling inefficiencies when employing such distributions.
However, there are several ways in which the sampling inefficiency could be a non-issue.

\begin{figure}
    \centering
    \includegraphics[width=0.8\linewidth]{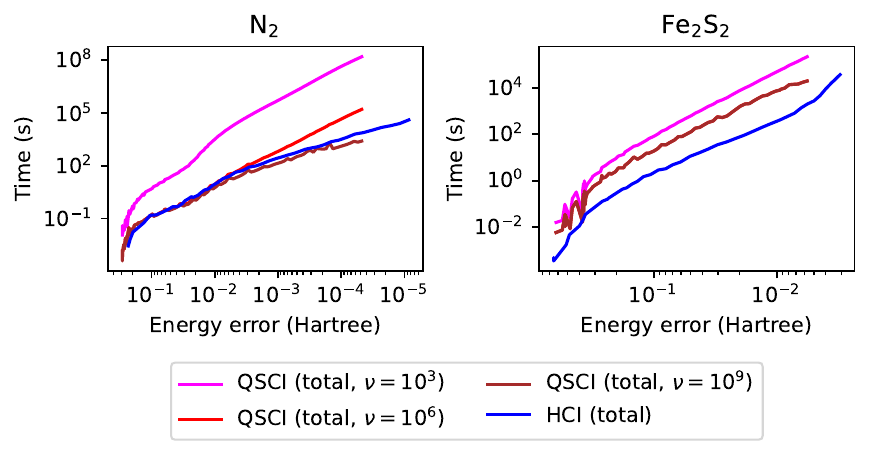}
    \caption{Total calculation time required to reach a given energy accuracy relative to the CASCI solution for HCI and QSCI on the N$_2$ (left) and [2Fe-2S] systems (right). The number of quantum samples (QSCI) and the threshold $\varepsilon$ are varied to adjust the energy accuracy. For HCI, iteration is stopped when the search step adds fewer than 1\% of the current number of determinants. The cumulative time (determinant search, Hamiltonian construction, and diagonalization) across all iterations is reported. For QSCI, the timings include a single diagonalization in addition to quantum sampling time, assuming the three indicated sampling rates $\nu$. For the iron-sulfur cluster, the cost of sampling becomes negligible with the higher sampling rates, causing the red $\nu=10^6$ and brown $\nu=10^9$ lines to overlap. }
    \label{fig:accuracy_scaling}
\end{figure}

First, if future quantum devices turn out to be orders of magnitude faster than current hardware, brute-force sampling from the ground-state distribution might be a viable strategy, although significant improvements to the performance of quantum hardware would be required.
In some of the cases presented in the Supporting Information (e.g., NH$_3$), reaching even a moderate accuracy requirement of $10^{-3}$ Hartree requires sampling counts of up to $10^9$.
In Ref. \citenum{robledo2024chemistry}, the authors report a sampling rate of three million executions for 45 minutes of QPU runtime, which translates to a sampling rate of around 1000 samples per second.
The work in Ref. \citenum{robledo2024chemistry} is based on superconducting qubit technology, which can likely be considered “state-of-the-art” in terms of execution speed.
With such a sample rate, performing $10^9$ quantum samples would take 10 days of continuous QPU execution for the sampling routine of a single calculation in an ideal, noise-free setting (excluding classic diagonalization costs). This may be considered prohibitively expensive, especially considering that the corresponding complete HCI calculations (including diagonalization costs) can be executed in about 15 minutes on a single CPU core. Nevertheless, the performance of future quantum devices remains uncertain. In Figure \ref{fig:accuracy_scaling}, we plot the accuracy (error relative to the CASCI reference) against the total calculation time for HCI and QSCI on the N$_2$ and [2Fe-2S] systems. For QSCI, we show results assuming three different sample-to-seconds conversion rates, $\nu$.
When assuming near-term sampling rates ($\nu=10^3$, magenta lines), QSCI is not competitive with HCI for either system. 
 The nitrogen molecule is an example where the quality of the selected determinants with QSCI is roughly equal to that of HCI (see Figure \ref{fig:n2_detail_energy}), but sampling efficiency in QSCI is poor.
 Thus, while the diagonalization costs are roughly comparable, with some advantage to QSCI, since only a single round of diagonalization is required, the quantum sampling component can be costly. 
 With faster sampling rates of $\nu=10^6$ (red lines), QSCI performs on par with HCI on the N$_2$ molecule up to an accuracy of about 5 milli-Hartree, beyond which the costs from quantum sampling make QSCI more expensive overall. Assuming even faster sampling rates of $\nu=10^9$ (brown lines), QSCI would outperform HCI, with total calculation time reduced by roughly a factor of three for the accuracy of $10^{-4}$ Hartree.
For the [2Fe-2S] system, sampling efficiency with QSCI is reasonably high, but the compactness of the resulting CI expansions is lower than the corresponding HCI expansions (see Figure \ref{fig:fe2s2_energy}). Thus, we find that for QSCI, the diagonalization costs become dominant. As a result, HCI remains more efficient from an accuracy-to-computation time standpoint throughout.
In summary, improvements to the execution speeds of future quantum hardware would have to be quite significant for the QSCI method to become cost-effective, and for some ground-state probability distributions, any advantage seems out of reach. Given that SQD is promoted as a current to near-term algorithm, the sampling efficiency is a critical problem.

Second, one could consider alternative sampling strategies.
For a fixed number of determinants ($L$), a very efficient sampling strategy would likely involve sampling from a uniform distribution over the $L$ largest CI coefficients, requiring approximately $O(2 L \log(L))$ samples to achieve near-full coverage\cite{robledo2024chemistry}. While we cannot rule out the existence of accurate and sample-efficient quantum circuits, their practical design and implementation are unknown and have yet to be demonstrated for QSCI. It constitutes a highly non-trivial problem and would most likely require a solution to the FCI problem in the first place to find the $L$ largest coefficients.
In contrast, sampling from the ground state distribution (or at least approximations thereof) seems feasible and has already been demonstrated to some extent.

\begin{figure}
    \centering
    \includegraphics[width=0.9\linewidth]{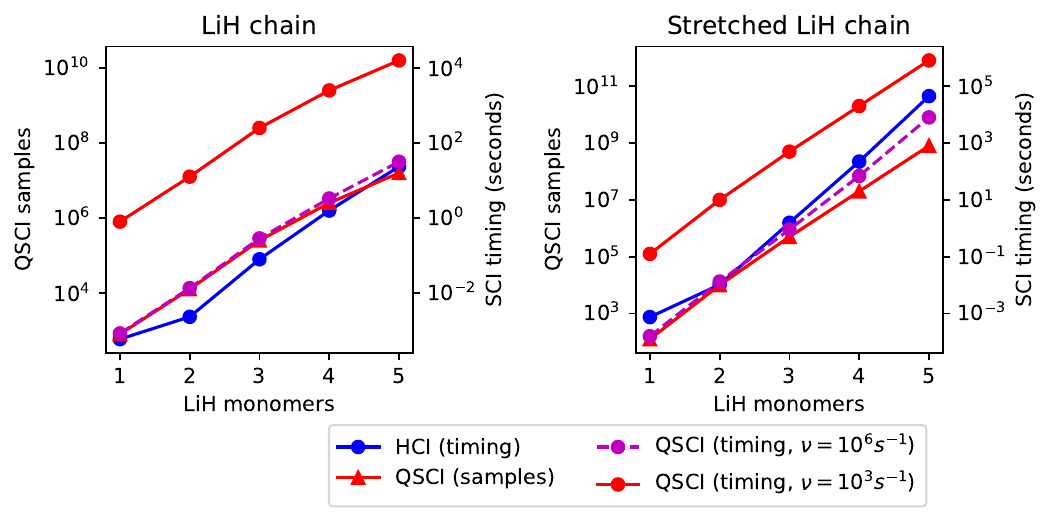}
    \caption{Cost scaling of QSCI and HCI (total calculation time on a single CPU core, seconds) for linear $(\mathrm{LiH})_n$ chains. For QSCI, we plot timings assuming either a near-term sampling rate $\nu$ of $10^3$ samples per second or a more optimistic estimate of $10^6$ samples per second. The plotted QSCI timings include the costs of Hamiltonian construction, diagonalization, and quantum sampling (with a conversion factor $1/\nu$). We consider two regimes: a weakly correlated regime (left) with $R_{\mathrm{Li-H}}=1.61$ \AA, $R_{\mathrm{LiH-LiH}}=1.81$ \AA, and a strongly correlated regime (right) with $R_{\mathrm{Li-H}}=3.22$ \AA, $R_{\mathrm{LiH-LiH}}=3.62$ \AA. A STO-3G basis was used, and a CAS($2n$e, $5n$o) active space was adopted (i.e., frozen-core). The shot count (for QSCI) or $\varepsilon$ (for HCI) was adjusted to reach a fixed $10^{-3}$ Hartree accuracy with respect to the CASCI reference energy.}
    \label{fig:LiH_chain_scaling}
\end{figure}

A further problem arises when scaling towards larger systems. As shown in Figure \ref{fig:LiH_chain_scaling}, both QSCI and HCI incur exponentially increasing computational costs with increasing system size, which comes as no surprise, since any SCI method (independent of the sampling being done classically or quantum) inherently has an exponential computational scaling with system size\cite{ammar2024compactification}. 
We use linear chains of lithium hydride as an example. The bond- and intermolecular distances are selected to probe either weakly (near-equilibrium) or strongly correlated (stretched) regimes.
The weakly correlated regime features a probability distribution that is similar to that of the previously discussed N$_2$, while the strongly correlated regime has a flatter probability distribution similar to [2Fe-2S].
On a positive note, we find the sampling costs from QSCI show an exponential scaling with a smaller exponent than the corresponding complete HCI calculation time. Thus, with high-enough sampling rates, QSCI could have the potential to show usefulness.

For the largest chain size considered, $(\mathrm{LiH})_5$, assuming a sampling rate of $10^3$, which is roughly representative of current capabilities\cite{robledo2024chemistry}, we find that QSCI is about 500 times slower in the weakly correlated regime, and 19 times slower in the strongly correlated regime.
Nevertheless, the slope (i.e, exponent) for QSCI at $\nu=10^3$ samples per second is lower than the HCI exponent, suggesting that a cross-over point should occur at some point, even when QSCI starts out slower than HCI. Extrapolation of the timings suggests that this will happen at $(\mathrm{LiH})_7$, but with a calculation cost of about 40 years.
If one instead assumes a sampling rate of $10^6$ samples per second (i.e., three orders of magnitude faster than current rates), QSCI becomes on par with HCI in terms of timing in the weakly correlated regime, and five times faster in the more strongly correlated regime.
Thus, without considerable improvements to the quantum sampling rates, it is difficult to consider QSCI a cost-effective method. 
Moreover, because QSCI is assessed under ideal conditions here, relying on hardware advances would ultimately require it to compete with fault-tolerant algorithms that avoid the exponential scaling.

\FloatBarrier
\section{Conclusion}\label{sec:conclusion}
In this work, we have explored the application of the Quantum Selected Configuration Interaction (QSCI) method in an ideal setting without device noise and assuming an optimized exact wave function ansatz is available. We investigated the performance and limitations of QSCI when applied to the nitrogen molecule (N$_2$) and the iron-sulfur cluster [2Fe-2S] when sampling from the ground-state distribution and identified two fundamental challenges that significantly impede the practicality of QSCI in realistic electronic structure calculations.

In the first case, exemplified by the nitrogen molecule, the QSCI method produces high-quality CI expansions comparable to those obtained through classical SCI heuristics. However, achieving high accuracy becomes increasingly difficult due to the problem of repeatedly sampling the same determinants. This inefficiency leads to difficulties in finding new determinants, making the process less effective as more samples are added. Since achieving near-FCI quality results (i.e., recovering the correlation energy) requires CI expansions with a large number of determinants, the high sampling demands of the QSCI method make it impractical for useful applications.

Alternatively, one may find systems such as the more strongly correlated iron-sulfur cluster [2Fe-2S], where sampling turns out to be less of a problem due to a flatter distribution of the CI coefficients. Unfortunately, the CI expansions generated by the QSCI method are much less compact than those generated by classical heuristics in these cases. HCI attains better energies with fewer determinants, leading to lower costs in the expensive diagonalization step compared to QSCI. This, again, limits the practical usefulness of QSCI.

Our investigation into scaled probability distributions further supports the idea that the sampling problem is fundamental to QSCI. We demonstrate that tuning the probability distribution can improve either the discovery of new determinants or the compactness of the CI expansion, but not both simultaneously. This trade-off highlights a key limitation of the QSCI approach. We also find that these problems significantly increase when using an approximated ansatz, such as LUCJ, for the sampling routine.

Investigations into the system-size scaling show glimmers of hope for QSCI. Although both QSCI and HCI suffer from an exponential scaling with the system size, the exponent in the QSCI sampling routine appears to be lower than the cumulative determinant search and diagonalization costs for HCI. Unfortunately, to reach any practical advantage, one has to assume extremely optimistic improvements to quantum sampling rates. 

For QSCI methods to have a practical advantage over classical SCI, clear gains must be demonstrated either in accuracy or efficiency. 
One possibility would be that QSCI could match or exceed the accuracy of classical SCI while keeping the cost of determinant selection low, giving better energies and wavefunctions without higher overhead. While we do find cases where the energy accuracy is comparable between QSCI and classical SCI, the sampling overheads for these cases are problematic.
Alternatively, QSCI could offer value if its sampling procedure is markedly faster than classical methods at identifying important determinants, even if the final energies are somewhat less accurate. In principle, rapid sampling could outweigh modest accuracy losses. However, with current quantum devices operating at limited sampling rates, this potential advantage remains largely hypothetical.
In regimes where classical SCI already yields accurate results at low cost, QSCI is unlikely to be competitive, regardless of quantum hardware improvements.

Overall, while QSCI can, in principle, be used to generate selected CI expansions, its practical utility is hindered by its reliance on large numbers of samples or its difficulty in producing compact wavefunctions. Hence, even with the perfect quantum settings assumed for QSCI in this work, it ultimately falls behind more effective classical SCI heuristics.

\begin{acknowledgement}
We acknowledge the financial support of the Novo Nordisk Foundation for the focused research project \textit{Hybrid Quantum Chemistry on Hybrid Quantum Computers} (HQC)$^2$, grant number NNFSA220080996.
Computations and simulations for the work described herein were supported by the DeiC National HPC Centre, SDU.
\end{acknowledgement}

\section*{Data availability statement}
The data that support the findings of this study are available from the corresponding author upon reasonable request.
Due to the large size of some datasets, specific arrangements may need to be made for data sharing. All details necessary for reproducing the results are provided in the manuscript and supplementary information.
Scripts and input files for the CASCI, HCI, and QSCI calculations are available at \url{https://github.com/peter-reinholdt/qsci-benchmarks}.

\begin{suppinfo}
Effect of probability density function scaling (Eq. \eqref{eq:pdf_scaling}), N$_2$ ground-state energy error as a function of the number of quantum samples, additional scaling factors for the N$_2$ system, timings analysis QSCI and HCI, and CI probability distribution plots for N$_2$ and [2Fe-2S] systems.
Examples of QSCI performance for 29 additional chemical systems.
\end{suppinfo}
\bibliography{main}
\end{document}

\section{Scaled probability distributions and alternative sampling wave functions}

\begin{figure}
    \centering
    \includegraphics[width=7cm]{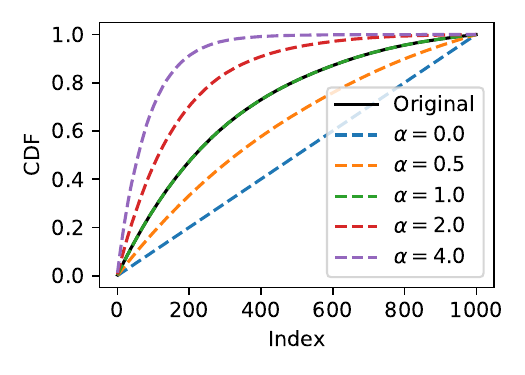}
    \caption{Effect of the probability distribution scaling. An input probability distribution function is scaled according to Eq. \eqref{eq:pdf_scaling}. We plot the original cumulative distribution function (CDF) (black line) and CDFs for scaled probability distributions with different parameter values $\alpha$.}
    \label{fig:pdf_scaling}
\end{figure}

We illustrate the effect of scaling in Figure \ref{fig:pdf_scaling} using a simple example probability distribution. The probability scaling is done according to 
\begin{equation}
    \tilde{p}_I = \frac{p_I^{\alpha}}{\left(\sum_J p_J^\alpha\right)}.
    \label{eq:pdf_scaling}
\end{equation}
The figure shows the cumulative distribution function of a (sorted) probability distribution function. Scaled versions of the same distribution are also shown.
As noted in the main manuscript, the original distribution of the determinants is recovered with $\alpha = 1$. Letting $\alpha>1$ emphasizes the larger weights in the probability distribution, while letting $\alpha<1$ emphasizes smaller weights in the probability distribution. At $\alpha=0$, this distribution corresponds to uniform sampling from the correct particle sector.

\begin{figure}
    \centering
    \includegraphics[width=0.7\linewidth]{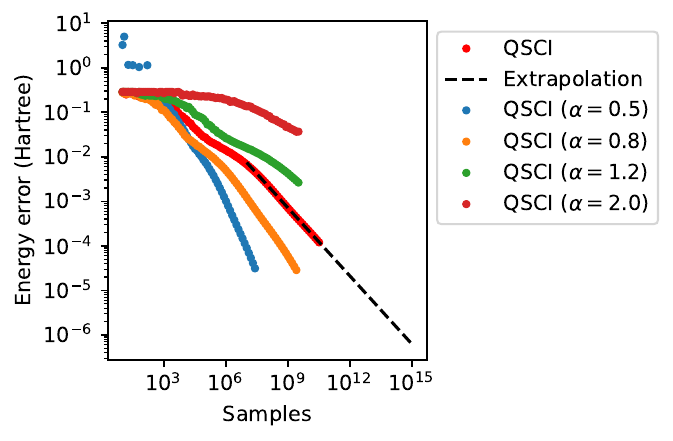}
    \caption{Error in the ground-state energy of the N$_2$/cc-pVDZ molecule ($R=1.09$ \AA) with a (10e,22o) active space, relative to the CASCI reference. The plot includes a power-law extrapolation of the QSCI data ($y=ax^b$, $a=27.99, b=-0.51, R^2=0.9996$) based on the points from $10^7$ samples and onwards. Data from the scaled CASCI distributions are also included.}
    \label{fig:n2_detail_samples}
\end{figure}

Figure \ref{fig:n2_detail_samples} shows the error in the energy as a function of the number of quantum samples. We include a power-law extrapolation of the energy error for the default QSCI method (sampling from the CASCI distribution), which shows that reaching micro-hartree precision would require around $10^{14}$ samples. 
We also include sampling from scaled distributions. With $\alpha<1$, we see that the energy error decreases more rapidly than sampling from the default CASCI distribution ($\alpha=1$), and conversely, letting $\alpha>1$ leads to slower convergence. We note that this figure only concerns the sample efficiency, but \emph{not} the compactness of the wave function as it plots energy against sample number, not against determinants found for the SCI expansion. For this, we refer to Figure 3 in the main text.

\begin{figure}
    \centering
    \includegraphics[width=1.0\linewidth]{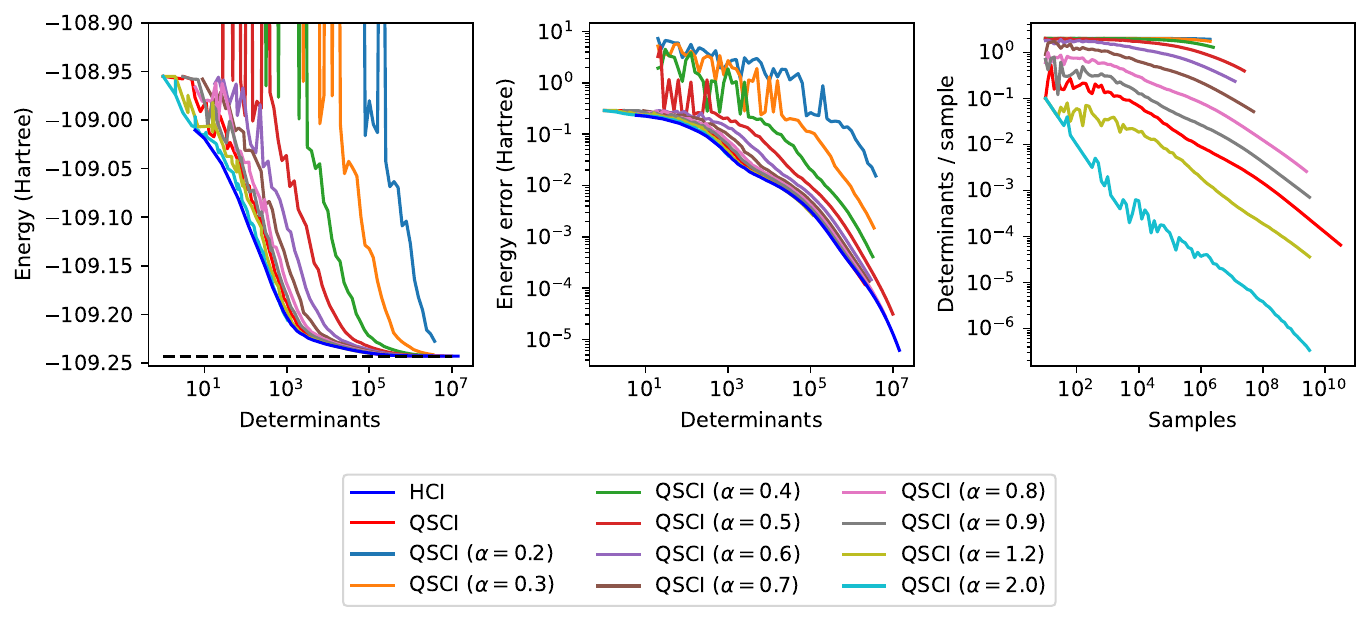}
    \caption{Ground-state energy of the N$_2$/cc-pVDZ molecule ($R=1.09$ \AA) with a (10e,22o) active space. The left panel shows total energies as a function of the number of determinants included in the CI expansion, with the black dashed lines indicating the CASCI energy.  
    The middle panel shows energy errors from HCI and QSCI wave functions (relative to the CASCI reference) as a function of the number of determinants included in the CI expansion. The right panel shows the determinants per sample for QSCI as a function of the number of samples. For QSCI, we include sampling from the original CASCI distribution, as well as scaled CASCI distributions.}
    \label{fig:n2_detail_energy_detailed_many}
\end{figure}

Figure \ref{fig:n2_detail_energy_detailed_many} is similar to Figure 3 from the main manuscript, but shows additional scaling factors $\alpha$. By adjusting the scaling factor, the sampling efficiency (determinants per sample) can systematically be increased, but at the cost of poorer CI expansion compactness (or vice-versa). 

\FloatBarrier
\section{Timing and scaling analysis}
In the following, we provide details about the scaling of QSCI and HCI when increasing the number of determinants (accuracy) in the CI expansion.
We compare the search costs between QSCI and HCI, defined as either the number of quantum samples (QSCI) or the determinant search costs (HCI).
Additionally, we note that since HCI is an iterative method, repeated diagonalization steps are required. Thus, we also provide an analysis of the cumulative computational costs for HCI, including determinant search, Hamiltonian construction, and Davidson diagonalization.

Table \ref{tab:hci_qsci_timings} reports computational timings for SCI diagonalization of the nitrogen molecule. We adjusted the number of samples (QSCI), or the threshold $\varepsilon$ (for HCI), to reach a target energy accuracy of about $10^{-3}$ or $10^{-4}$ Hartree relative to the FCI reference.
We also include sampling from a scaled probability distribution $(\alpha=0.5$) to gauge how the classical overhead from diagonalization increases in a situation where sampling is more efficient, but more non-important determinants become included in the CI expansion.
Since HCI is an iterative method, repeated diagonalizations are required, and we report the sum of timings across all iterations. For QSCI, only a single diagonalization step is, in principle, needed in an ideal setting without configuration recovery. Accordingly, for a similar number of determinants, QSCI diagonalization costs are lower than for HCI. 
To assign timings for the search component in QSCI, one has to decide on a conversion factor from quantum samples to seconds. As discussed in the main text, current/near-term sample rates for QSCI sampling on superconducting qubits appear to be around $\nu=10^3$ samples per second, but we also apply a more optimistic estimate of $\nu=10^6$ samples per second in Table \ref{tab:hci_qsci_timings}. Assuming current/near-term sampling rates ($\nu=10^3$ samples per second), the total computational cost for QSCI is significantly larger than that of HCI, regardless of the scaling factor. Under the future-leaning optimistic rate assumption ($\nu=10^6$ samples per second), for the scaled probability distributions ($\alpha=0.5$), the time associated with quantum sampling is overall negligible, adding only a few seconds to the total timing. However, when sampling from the original, unscaled ground-state distributions, the total cost becomes entirely dominated by the costs of quantum sampling. In particular, for the higher-accuracy ($10^{-4}$ Hartree), we find that QSCI becomes more than ten times slower than the corresponding classical HCI calculations.

\begin{table}
    \centering
\begin{tabular}{|l|rrrrrrr|}
\hline 
 & Samples & Determinants & Accuracy (Ha) & $t_{search}$ (s) & $t_{ham}$ (s) & $t_{diag}$ (s) & $t_{total}$ (s)\tabularnewline
\hline 
HCI & - & 322929 & $0.963\times10^{-3}$ & 140.3 & 147.5 & 93.5 & 381.3\tabularnewline
HCI$^{a}$ & - & 322929 & $0.963\times10^{-3}$ & 80.7 & 147.5 & 60.0 & 288.2\tabularnewline
QSCI & $5\times10^{8}$ & 322234 & $1.02\times10^{-3}$ & - & 142.3 & 7.7 & 150.0\tabularnewline
QSCI$^b$ & $5\times10^{8}$ & 322234 & $1.02\times10^{-3}$ & 500000 & 142.3 & 7.7 & 500000\tabularnewline
QSCI$^c$ & $5\times10^{8}$ & 322234 & $1.02\times10^{-3}$ & 500 & 142.3 & 7.7 & 650.0\tabularnewline
QSCI\phantom{$^b$} $(\alpha=0.5)$ & $10^{6}$ & 989062 & $1.00\times10^{-3}$ & - & 487.2 & 28.2 & 515.4\tabularnewline
QSCI$^b$ $(\alpha=0.5)$ & $10^{6}$ & 989062 & $1.00\times10^{-3}$ & 1000 & 487.2 & 28.2 & 1515.4\tabularnewline
QSCI$^c$ $(\alpha=0.5)$ & $10^{6}$ & 989062 & $1.00\times10^{-3}$ & 1 & 487.2 & 28.2 & 516.4\tabularnewline
\hline 
 & Samples & Determinants & Accuracy (Ha) & $t_{search}$ (s) & $t_{ham}$ (s) & $t_{diag}$ (s) & $t_{total}$ (s)\tabularnewline
\hline 
HCI & - & 2667952 & $0.948\times10^{-4}$ & 934.8 & 1677.9 & 2106.4 & 4719.2\tabularnewline
HCI$^{a}$ & - & 2667888 & $0.948\times10^{-4}$ & 514.7 & 1677.9 & 1264.4 & 3457.0\tabularnewline
QSCI & $5\times10^{10}$ & 2529629 & $0.932\times10^{-4}$ & - & 1498.6 & 236.2 & 1734.7\tabularnewline
QSCI$^b$ & $5\times10^{10}$ & 2529629 & $0.932\times10^{-4}$ & 50000000 & 1498.6 & 236.2 & 50000000\tabularnewline
QSCI$^c$ & $5\times10^{10}$ & 2529629 & $0.932\times10^{-4}$ & 50000 & 1498.6 & 236.2 & 51734.7\tabularnewline
QSCI\phantom{$^b$} $(\alpha=0.5)$ & $10^{7}$ & 5434904 & $0.900\times10^{-4}$ & - & 3494.5 & 449.8 & 3944.3\tabularnewline
QSCI$^b$ $(\alpha=0.5)$ & $10^{7}$ & 5434904 & $0.900\times10^{-4}$ & 10000 & 3494.5 & 449.8 & 13944.3\tabularnewline
QSCI$^c$ $(\alpha=0.5)$ & $10^{7}$ & 5434904 & $0.900\times10^{-4}$ & 10 & 3494.5 & 449.8 & 3954.3\tabularnewline
\hline 
\end{tabular}
    \caption{Computational timings for SCI diagonalization on a single CPU core for  N$_2$/cc-pVDZ molecule ($R=1.09$ \AA) with a (10e,22o) active space. Cost include determinant search $t_{search}$, sparse hamiltoninan construction $t_{ham}$, and Davidson diagonalization $t_{diag}$. For HCI, the sum of timings across all iterations is reported. The total time is defined $t_{total} = t_{search} + t_{ham} + t_{diag}$ for HCI and $t_{total} = t_{ham} + t_{diag}$ for QSCI. We assign a conversion factor from quantum samples to $t_{search}$ in the rows marked $^b$ and $^c$. \\$^a$HCI iteration is stopped when the search step adds fewer than 1\% of the current number of determinants. 
    \\$^b$Assuming a quantum sampling rate of $\nu=10^3s^{-1}$.\\$^c$Assuming an optimistic future quantum sampling rate of $\nu=10^6s^{-1}$.} 
    \label{tab:hci_qsci_timings}
\end{table}

\begin{figure}
    \centering
    \includegraphics[width=1.0\linewidth]{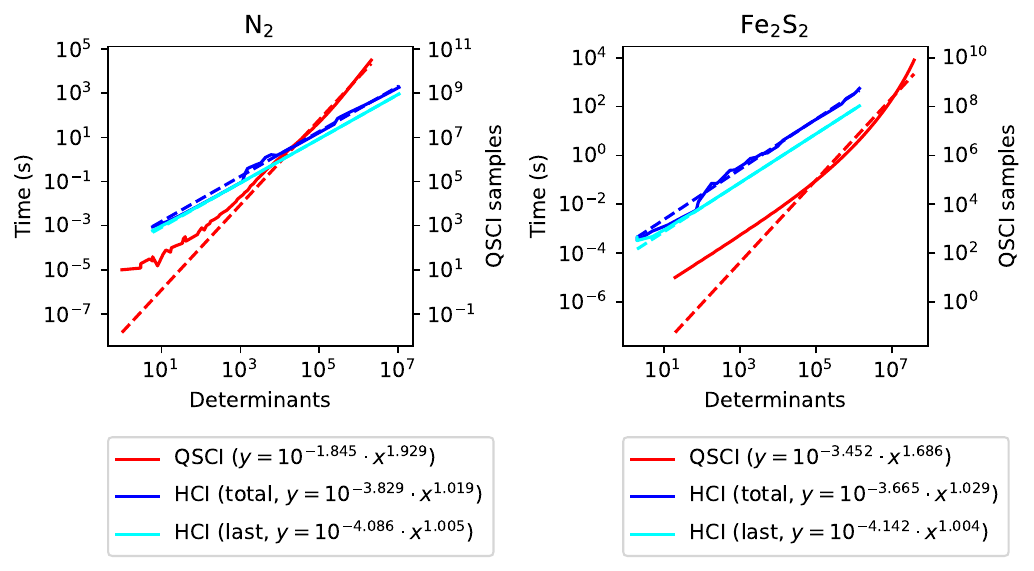}
    \caption{Determinant search cost scaling for HCI and QSCI for the N$_2$/cc-pVDZ molecule ($R=1.09$ \AA) with a (10e,22o) active space (left) and the [2Fe-2s] system (30e,20o) (right). For HCI, we report the total timing across all iterations (blue lines) and the time required for a single iteration (cyan lines). For QSCI (red lines), we plot the number of determinants found for a given number of shots. The two $y$-scales are aligned assuming a quantum sampling rate of $10^6$ samples per second. Power law fits ($y=ax^b$) are shown as dashed lines. Note that the fit for the final search iteration in HCI (cyan dashed lines) overlaps almost completely with the data (cyan lines). HCI iteration is stopped when the search step adds fewer than 1\% of the current number of determinants.}
    \label{fig:search_cost}
\end{figure}

\begin{figure}
    \centering
    \includegraphics[width=1.0\linewidth]{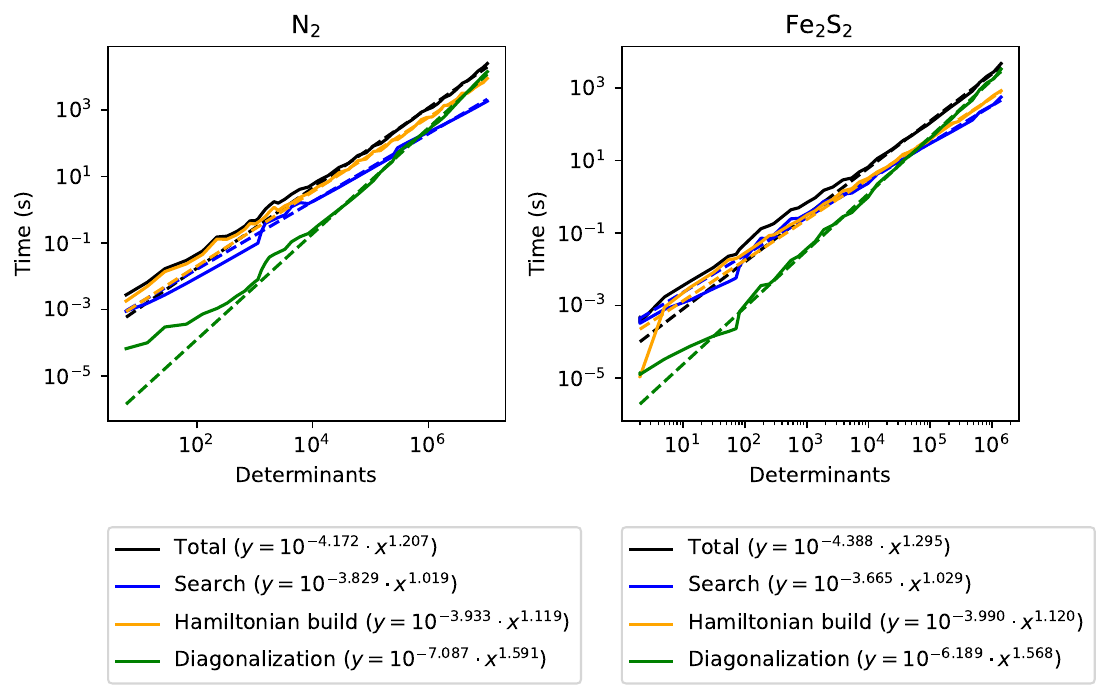}
    \caption{Cost scaling for HCI for the N$_2$/cc-pVDZ molecule ($R=1.09$ \AA) with a (10e,22o) active space. The threshold $\varepsilon$ is varied to generate different final number of determinants. Total timings and timing components for determinant search, sparse Hamiltonian construction, and Davidson diagonalization are shown. The sum of timings across all HCI iterations is shown in all cases. Power-law fits $y=ax^b$ are shown as dashed lines, fitted from the data with 10000 determinants and onwards. HCI iteration is stopped when the search step adds fewer than 1\% of the current number of determinants.}
    \label{fig:HCI_cost_scaling}
\end{figure}

Figure \ref{fig:search_cost} shows the determinant search cost for HCI and QSCI, defined as the number of shots (QSCI) or CPU time (HCI) required to reach a certain number of determinants. We present data collected by varying the number of shots or the threshold $\varepsilon$. Since HCI is an iterative method, multiple search iterations are involved as the CI expansion grows, and we report both total timings (blue line) and timings for the final search iteration (cyan line). The data is fitted to a power-law functional form $y=ax^b$ on the data points from 10000 determinants and onwards, which reveals roughly linear cost scaling for HCI (dashed blue line) and nearly quadratic cost scaling for QSCI with N$_2$ (dashed red line).
This analysis directly relates to the repeated sampling problem presented in the main text (e.g.,  Figure 2 for N$_2$ and Figure 4 for [2Fe-2S]). With QSCI, initially, there are relatively few instances of repeated sampling of the same determinants (high DPS), which gives a mostly linear scaling in the search cost. Only with larger CI expansions, the repeated sampling of the same determinants begins to become problematic (lower DPS), which is clear here as the sampling costs for QSCI curve upwards, giving steeper exponents as the size of the CI expansion is increased. Fitting from the last twenty data points gives apparent scalings of $x^{2.2}$ (N$_2$) or $x^{2.9}$ ([2Fe-2S]). 
Since the repeated sampling problem is less pronounced in [2Fe-2S] than in N$_2$, the scaling is influenced accordingly.
We note that the resource requirements in Figure \ref{fig:search_cost} only relate to the determinant selection routine, but that HCI (as an iterative process) includes additional overhead from Hamiltonian construction and diagonalization (see Figure \ref{fig:HCI_cost_scaling}).

Figure \ref{fig:HCI_cost_scaling} shows the total timings as well as individual timing components for HCI calculations on the nitrogen molecule and [2Fe-2S] with an increasingly tight threshold $\varepsilon$. In the beginning, the sparse Hamiltonian and search components dominate the cost, both of which scale nearly linearly with the number of determinants. Diagonalization eventually becomes the most significant cost component, due to a higher scaling exponent of around 1.6 (i.e., super-linear, but sub-quadratic scaling). 

\FloatBarrier
\section{CI vector plots}

\begin{figure}
    \centering
    \includegraphics[width=0.4\linewidth]{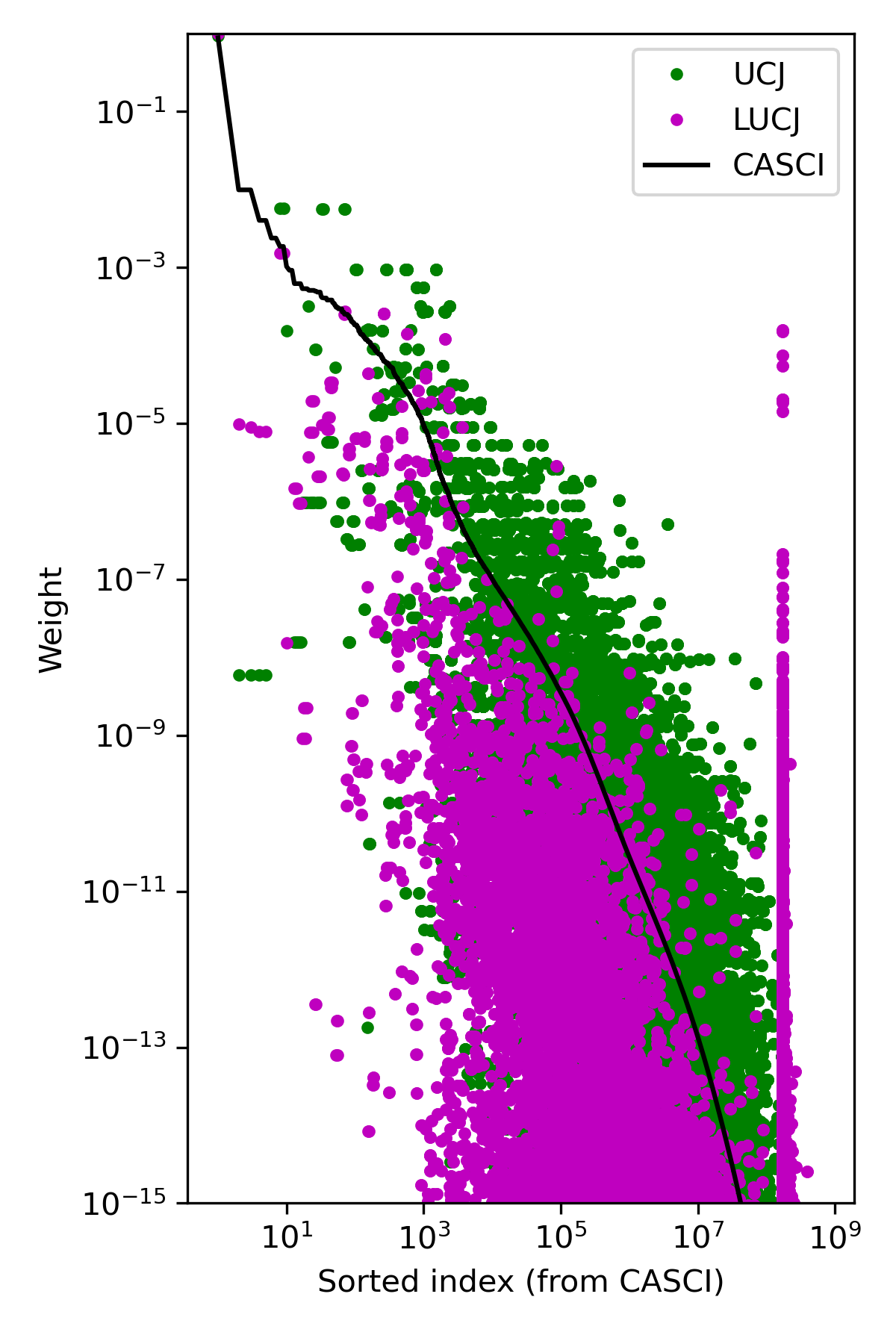}
    \caption{CI weights ($|c_I|^2$) of the N$_2$/cc-pVDZ molecule ($R=1.09$ \AA) with a (10e,22o) active space. The determinant weights are sorted in descending order according to the weights from the CASCI wave function.}
    \label{fig:ci_vector}
\end{figure}

Figure \ref{fig:ci_vector} plots the CI weights sorted in descending order according to the CASCI weights for N$_2$/cc-pVDZ in a (10e,22o) active space at $R$ = 1.09 Å. 
For UCJ and LUCJ, points appearing to the left of the solid black line (CASCI) correspond to under-estimated weights, while points appearing to the right of the solid black line correspond to over-estimated weights.    
Crucially, the correct ranking of the CI weights is clearly not preserved with UCJ or LUCJ.

\begin{figure}
    \centering
    \includegraphics[width=8cm]{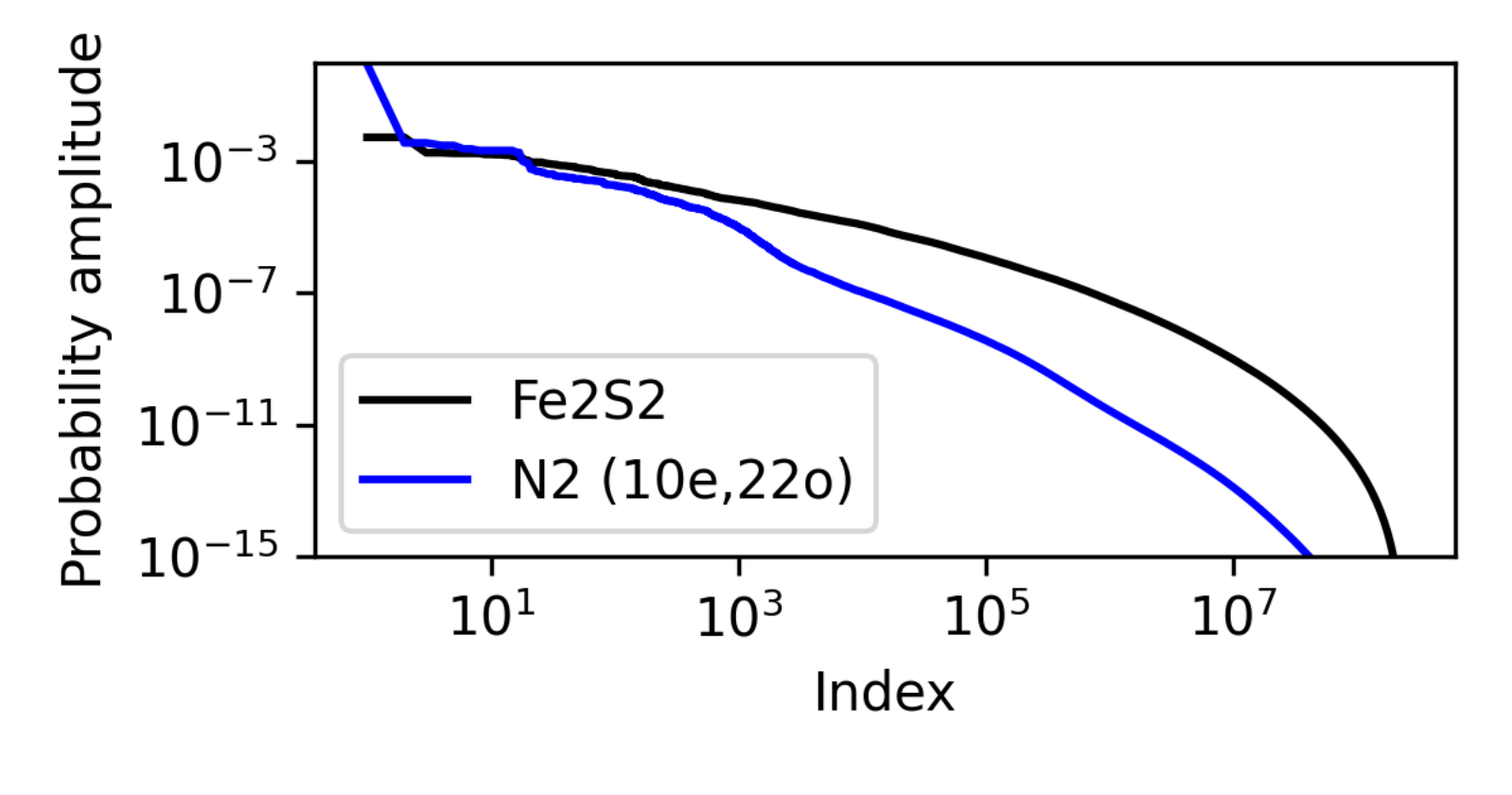}
    \caption{Sorted (descending order) CI probability amplitudes, $|c_I|^2$, for the CASCI wave functions of the [2Fe-2S] and N$_2$/cc-pVDZ  (10e,22o) systems.}
    \label{fig:fe2s2_civec}
\end{figure}

Figure \ref{fig:fe2s2_civec} plots the CI weights sorted in descending order for the [2Fe-2S] and N$_2$ systems. 
The [2Fe-2S] CASCI wave function amplitudes have a rather flat and slowly decaying distribution with no single large weight, as compared to the CASCI probability amplitudes of N$_2$.

\FloatBarrier
\section{Additional systems}
In the following, we provide further comparisons of the compactness and sample efficiency of QSCI for additional molecular systems.
We include two sets of systems, the first using geometries from the W4-17-MR benchmark set\cite{karton2017w4}, which is composed of systems with some multireference character, and a second set of hand-picked structures.
For the second set of structures, geometries were optimized with DFT (wB97X-D3BJ\cite{najibi2018nonlocal} with a def2-TZVP\cite{weigend2005balanced} basis set) in Orca\cite{neese2022software}, version 6.0.0, unless otherwise noted.
The twisted ethene geometry (with a 90-degree torsion) was obtained by running a geometry optimization with the H-C=C-H dihedrals constrained to $\pm$90 degrees.
The transition state geometry for the Diels-Alder reaction between furan and ethene  was taken from Ref. \citenum{kjellgren2024variational}, and is based on nudged-elastic band\cite{henkelman2000climbing,goumans2009embedded} calculations using CAM-B3LYP\cite{yanai2004new}-
D3BJ\cite{grimme2010consistent,grimme2011effect}/6-31G*\cite{ditchfield1971self}.

For each system, Hartree-Fock and CASCI wave functions, and one- and two-electron MO integrals (cc-pVDZ\cite{dunning1989gaussian} basis) were obtained using PySCF \cite{pyscf}.
The HCI and QSCI calculations were carried out as described in the main text.
\newpage
\subsection{W4-17-MR}

\begin{figure}
\centering
\includegraphics[width=16.0cm]{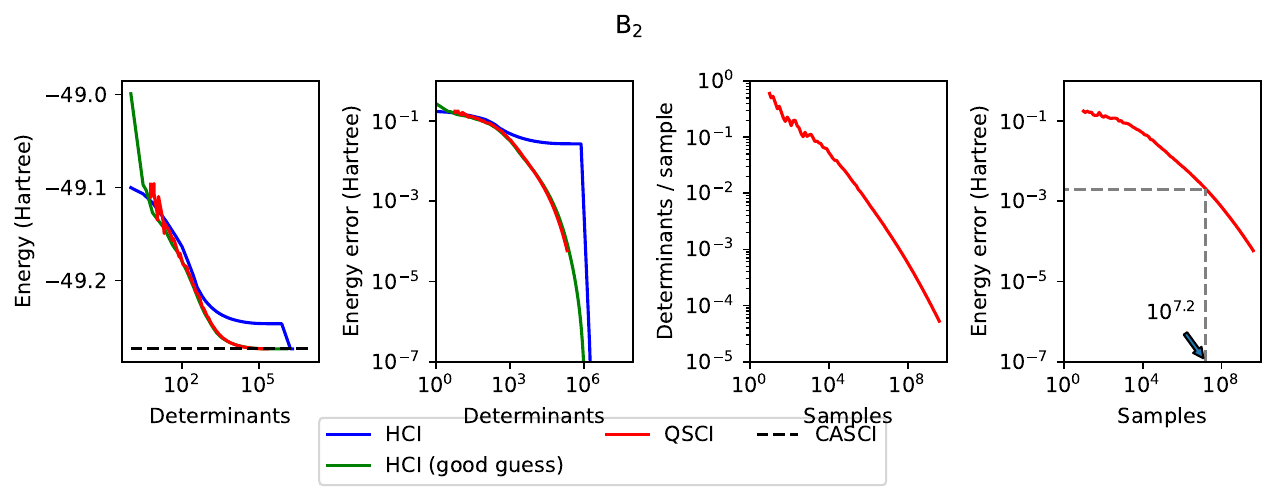}
    \caption{Ground state energy of the triplet B$_2$ molecule in an (6e,26o) active space.
    The two left panels show the energy (linear scale) and energy error relative to the CASCI reference (log scale).
    The panels to the right show the determinants per sample and the energy error (log scale) as a function of the number of samples. 
    The small arrow on the rightmost panel indicates the number of samples required to reach milli-Hartree precision. The "good" guess (green lines) for HCI uses a singly-excited SOMO-1-LUMO determinant as the initial CI expansion.}
\end{figure}

\begin{figure}
\centering
\includegraphics[width=16.0cm]{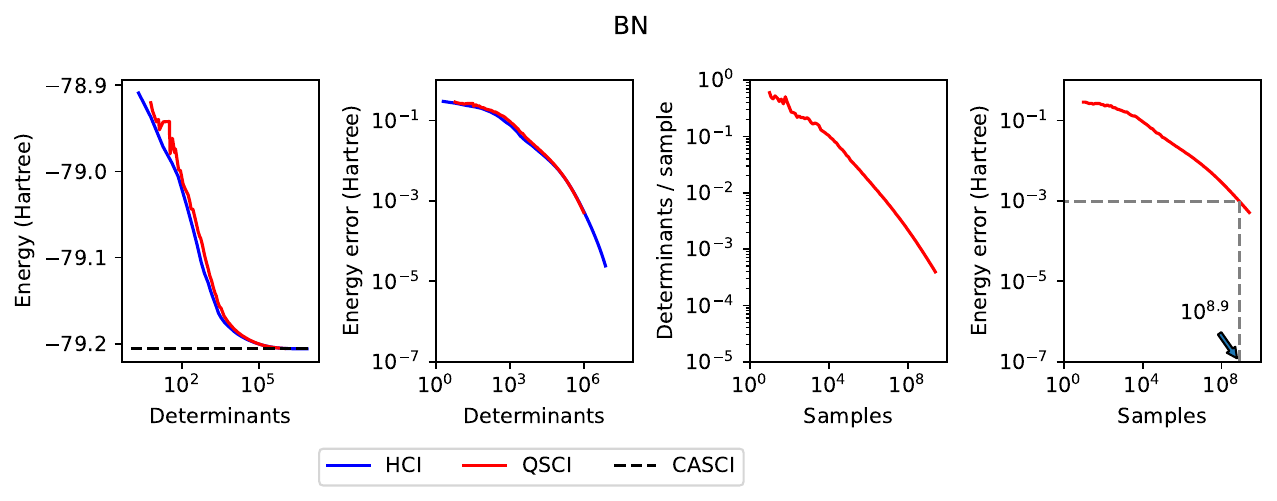}
    \caption{Ground state energy of the BN molecule in an (8e,26o) active space.
    The two left panels show the energy (linear scale) and energy error relative to the CASCI reference (log scale).
    The panels to the right show the determinants per sample and the energy error (log scale) as a function of the number of samples. 
    The small arrow on the rightmost panel indicates the number of samples required to reach milli-Hartree precision. }
\end{figure}

\begin{figure}
\centering
\includegraphics[width=16.0cm]{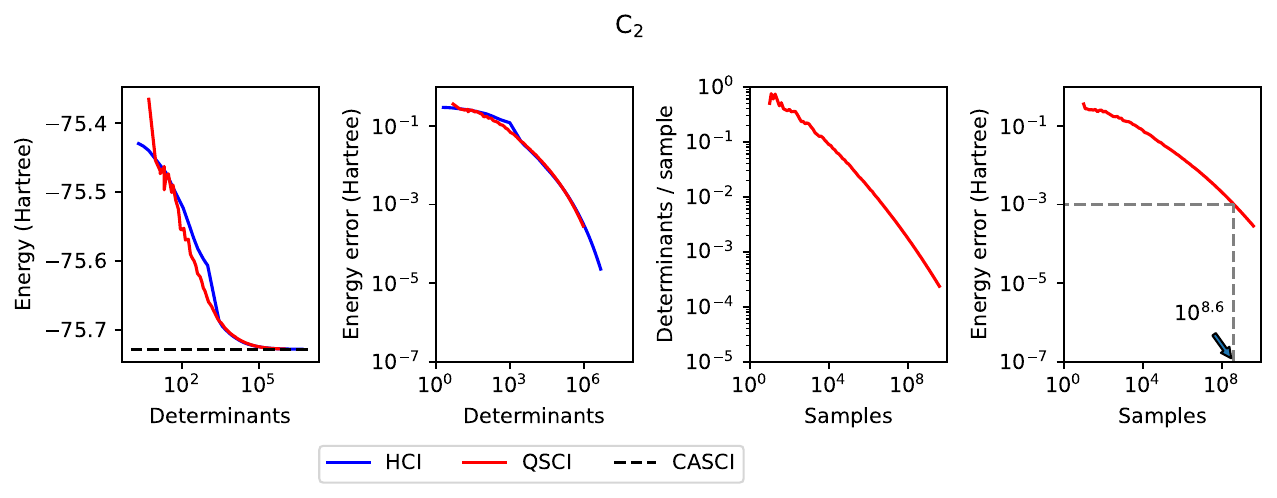}
    \caption{Ground state energy of the C$_2$ molecule in an (8e,26o) active space.
    The two left panels show the energy (linear scale) and energy error relative to the CASCI reference (log scale).
    The panels to the right show the determinants per sample and the energy error (log scale) as a function of the number of samples. 
    The small arrow on the rightmost panel indicates the number of samples required to reach milli-Hartree precision. }
\end{figure}

\begin{figure}
\centering
\includegraphics[width=16.0cm]{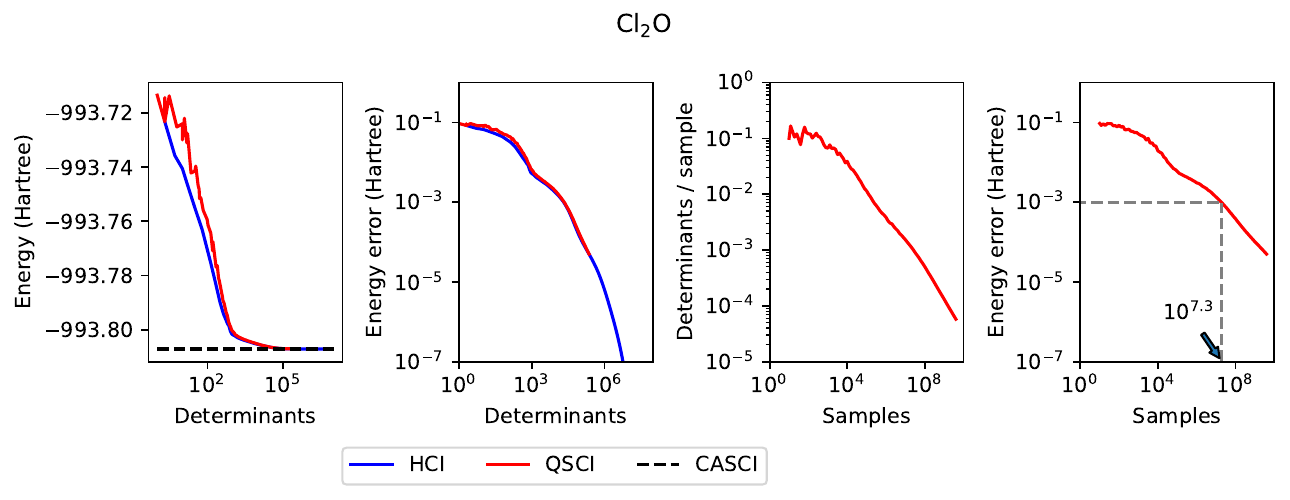}
    \caption{Ground state energy of the Cl$_2$O molecule in an (14e,16o) active space.
    The two left panels show the energy (linear scale) and energy error relative to the CASCI reference (log scale).
    The panels to the right show the determinants per sample and the energy error (log scale) as a function of the number of samples. 
    The small arrow on the rightmost panel indicates the number of samples required to reach milli-Hartree precision. }
\end{figure}

\begin{figure}
\centering
\includegraphics[width=16.0cm]{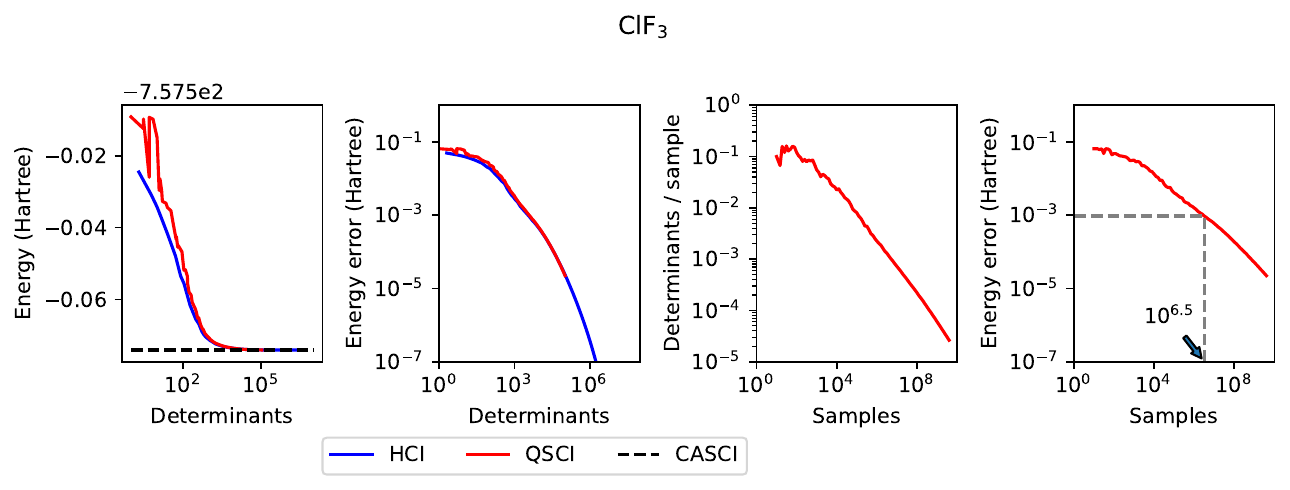}
    \caption{Ground state energy of the ClF$_3$ molecule in an (14e,16o) active space.
    The two left panels show the energy (linear scale) and energy error relative to the CASCI reference (log scale).
    The panels to the right show the determinants per sample and the energy error (log scale) as a function of the number of samples. 
    The small arrow on the rightmost panel indicates the number of samples required to reach milli-Hartree precision. }
\end{figure}

\begin{figure}
\centering
\includegraphics[width=16.0cm]{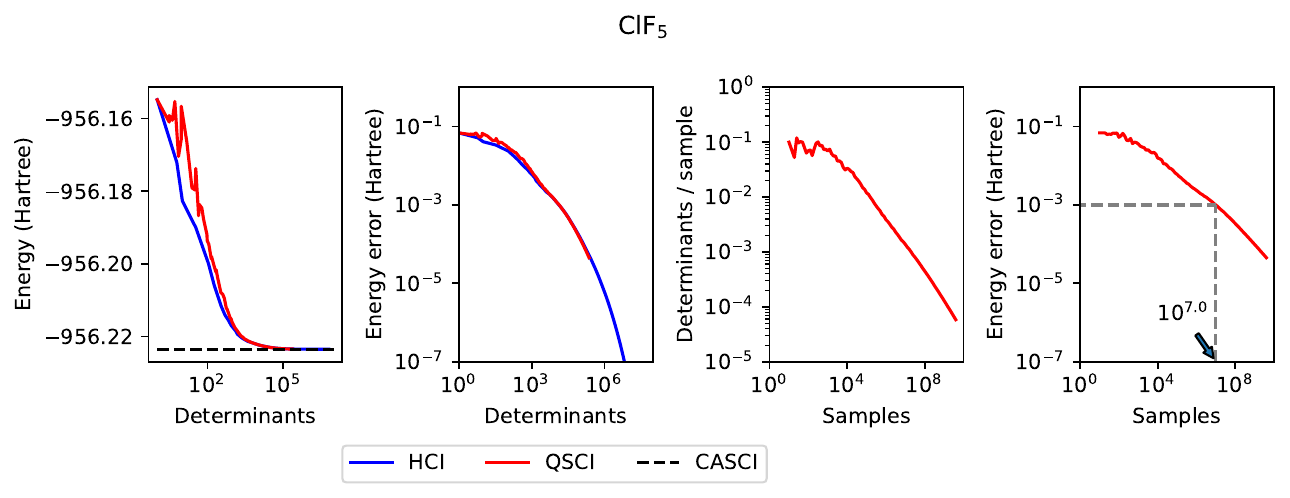}
    \caption{Ground state energy of the ClF$_5$ molecule in an (18e,16o) active space.
    The two left panels show the energy (linear scale) and energy error relative to the CASCI reference (log scale).
    The panels to the right show the determinants per sample and the energy error (log scale) as a function of the number of samples. 
    The small arrow on the rightmost panel indicates the number of samples required to reach milli-Hartree precision. }
\end{figure}

\begin{figure}
\centering
\includegraphics[width=16.0cm]{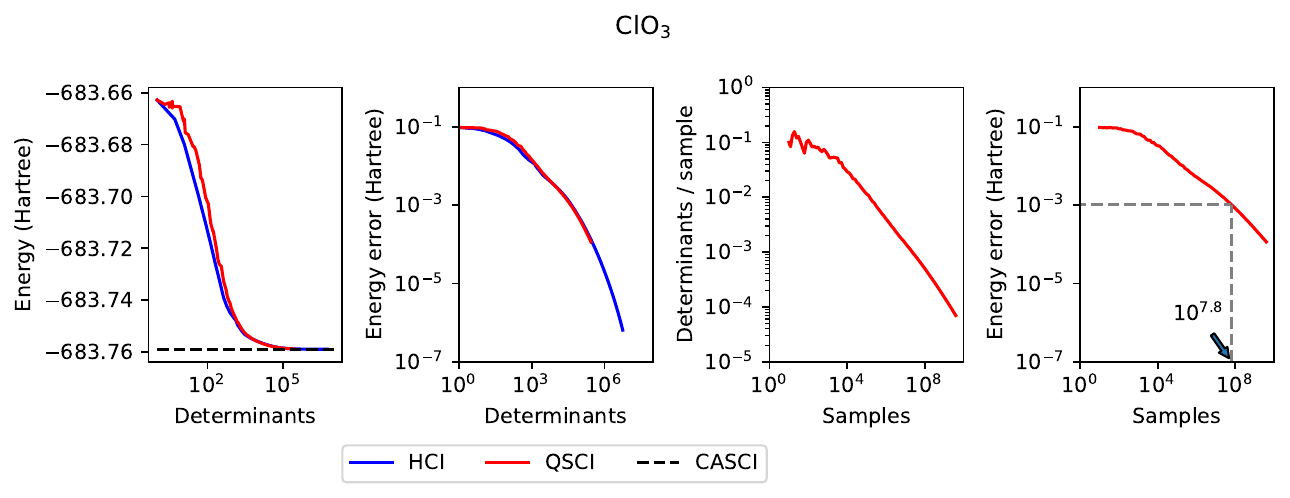}
    \caption{Ground state energy of the ClO$_3$ radical (doublet) in an (11e,18o) active space.
    The two left panels show the energy (linear scale) and energy error relative to the CASCI reference (log scale).
    The panels to the right show the determinants per sample and the energy error (log scale) as a function of the number of samples. 
    The small arrow on the rightmost panel indicates the number of samples required to reach milli-Hartree precision. }
\end{figure}

\begin{figure}
\centering
\includegraphics[width=16.0cm]{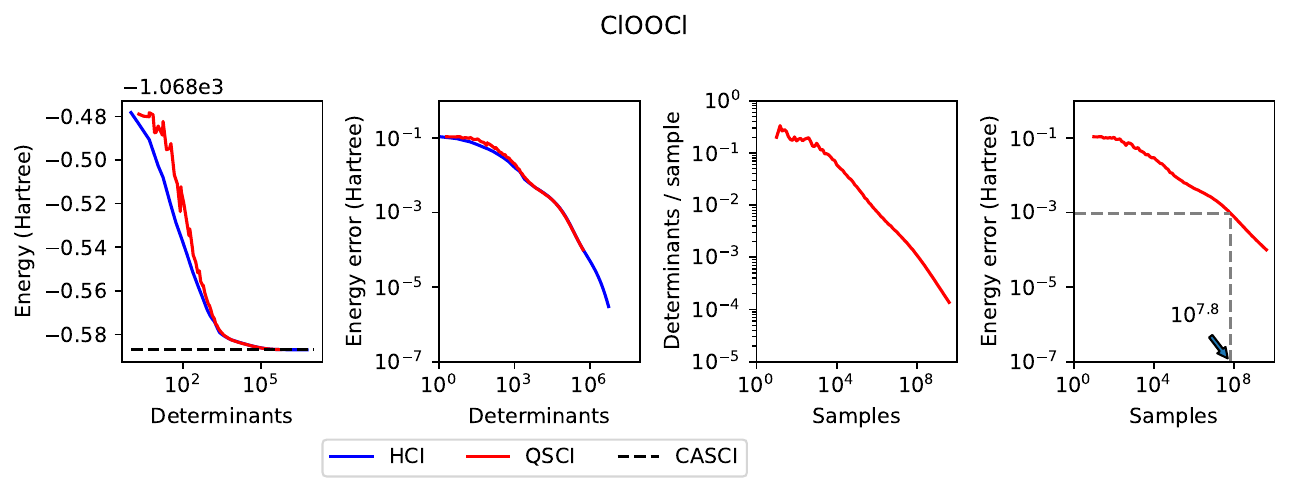}
    \caption{Ground state energy of the ClOOCl molecule in an (18e,16o) active space.
    The two left panels show the energy (linear scale) and energy error relative to the CASCI reference (log scale).
    The panels to the right show the determinants per sample and the energy error (log scale) as a function of the number of samples. 
    The small arrow on the rightmost panel indicates the number of samples required to reach milli-Hartree precision. }
\end{figure}

\begin{figure}
\centering
\includegraphics[width=16.0cm]{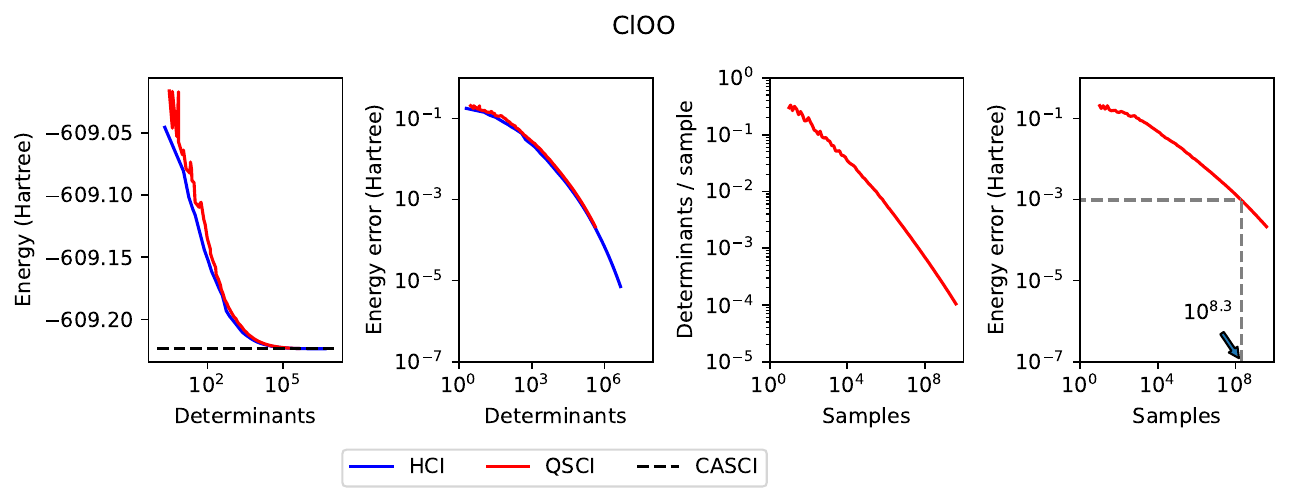}
    \caption{Ground state energy of the ClOO radical (doublet) in an (13e,18o) active space.
    The two left panels show the energy (linear scale) and energy error relative to the CASCI reference (log scale).
    The panels to the right show the determinants per sample and the energy error (log scale) as a function of the number of samples. 
    The small arrow on the rightmost panel indicates the number of samples required to reach milli-Hartree precision. }
\end{figure}

\begin{figure}
\centering
\includegraphics[width=16.0cm]{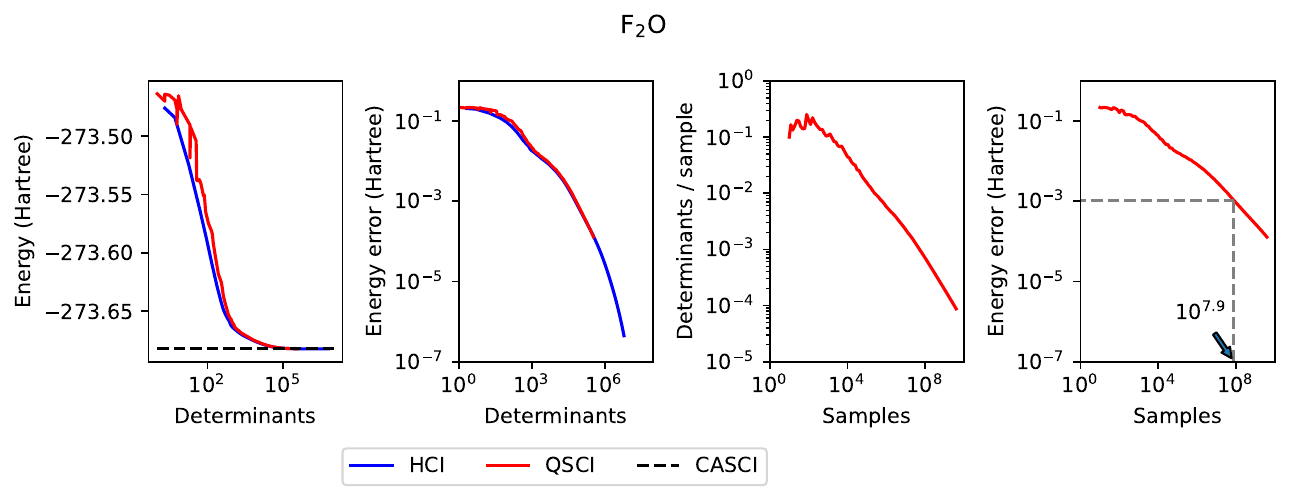}
    \caption{Ground state energy of the F$_2$O molecule in an (14e,16o) active space.
    The two left panels show the energy (linear scale) and energy error relative to the CASCI reference (log scale).
    The panels to the right show the determinants per sample and the energy error (log scale) as a function of the number of samples. 
    The small arrow on the rightmost panel indicates the number of samples required to reach milli-Hartree precision. }
\end{figure}

\begin{figure}
\centering
\includegraphics[width=16.0cm]{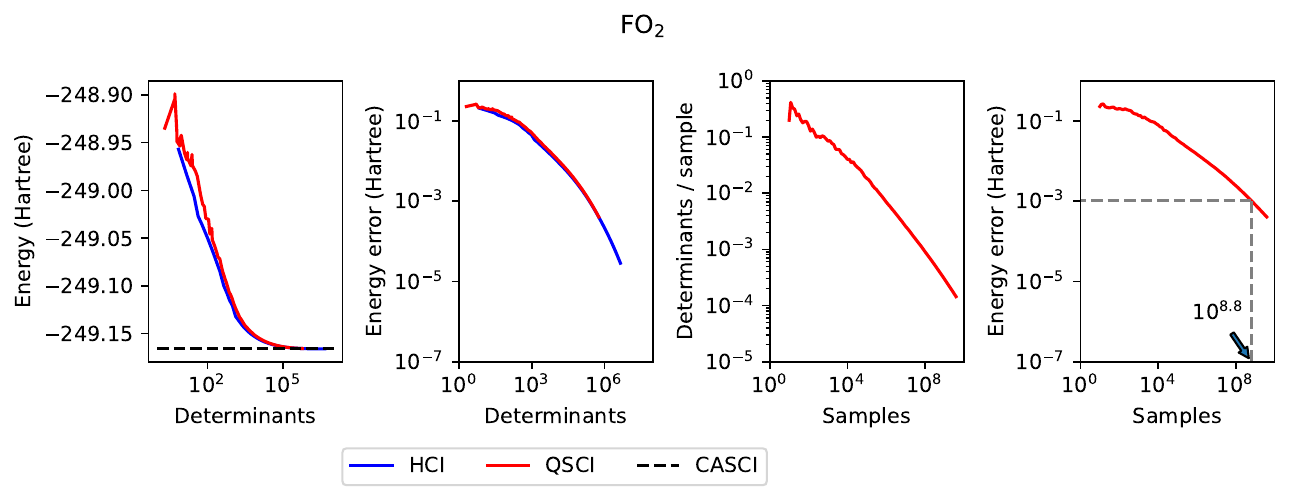}
    \caption{Ground state energy of the FO$_2$ radical (doublet) in an (13e,18o) active space.
    The two left panels show the energy (linear scale) and energy error relative to the CASCI reference (log scale).
    The panels to the right show the determinants per sample and the energy error (log scale) as a function of the number of samples. 
    The small arrow on the rightmost panel indicates the number of samples required to reach milli-Hartree precision. }
\end{figure}

\begin{figure}
\centering
\includegraphics[width=16.0cm]{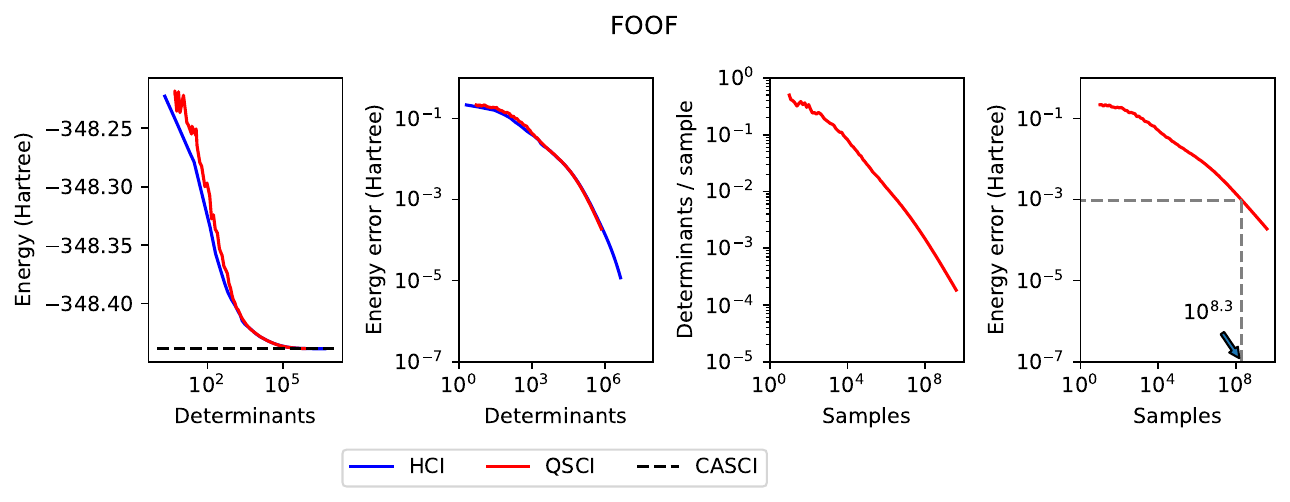}
    \caption{Ground state energy of the FOOF molecule in an (18e,16o) active space.
    The two left panels show the energy (linear scale) and energy error relative to the CASCI reference (log scale).
    The panels to the right show the determinants per sample and the energy error (log scale) as a function of the number of samples. 
    The small arrow on the rightmost panel indicates the number of samples required to reach milli-Hartree precision. }
\end{figure}

\begin{figure}
\centering
\includegraphics[width=16.0cm]{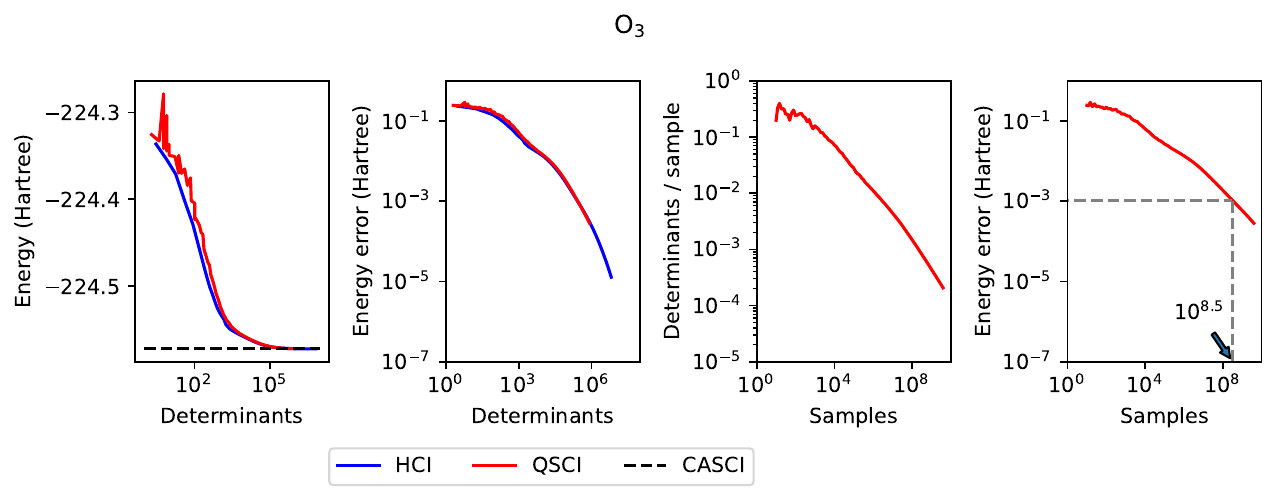}
    \caption{Ground state energy of the O$_3$ molecule in an (12e,18o) active space.
    The two left panels show the energy (linear scale) and energy error relative to the CASCI reference (log scale).
    The panels to the right show the determinants per sample and the energy error (log scale) as a function of the number of samples. 
    The small arrow on the rightmost panel indicates the number of samples required to reach milli-Hartree precision. }
\end{figure}

\begin{figure}
\centering
\includegraphics[width=16.0cm]{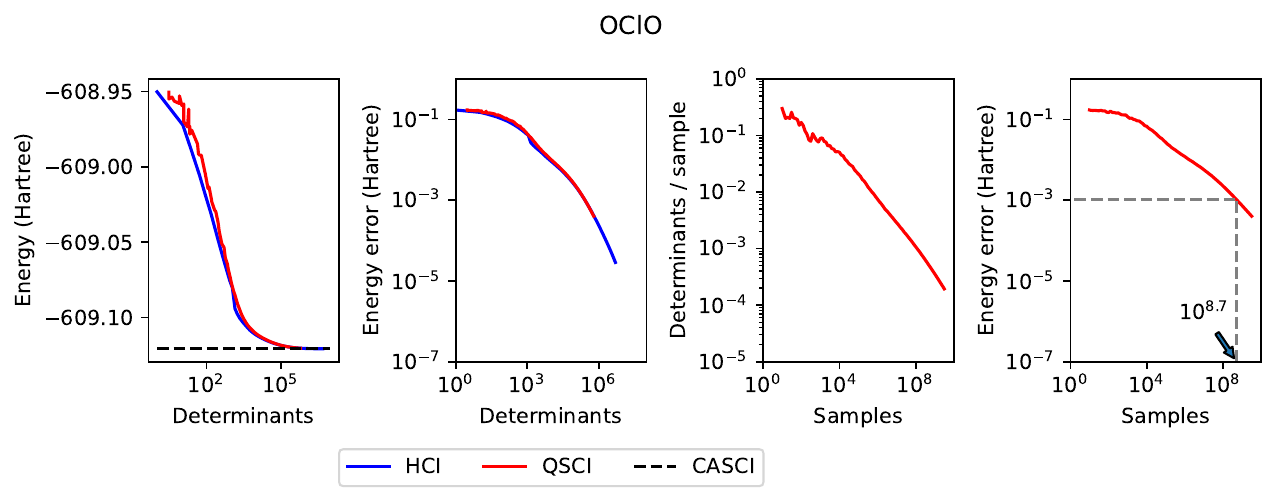}
    \caption{Ground state energy of the OClO radical (doublet) in an (13e,18o) active space.
    The two left panels show the energy (linear scale) and energy error relative to the CASCI reference (log scale).
    The panels to the right show the determinants per sample and the energy error (log scale) as a function of the number of samples. 
    The small arrow on the rightmost panel indicates the number of samples required to reach milli-Hartree precision. }
\end{figure}

\begin{figure}
\centering
\includegraphics[width=16.0cm]{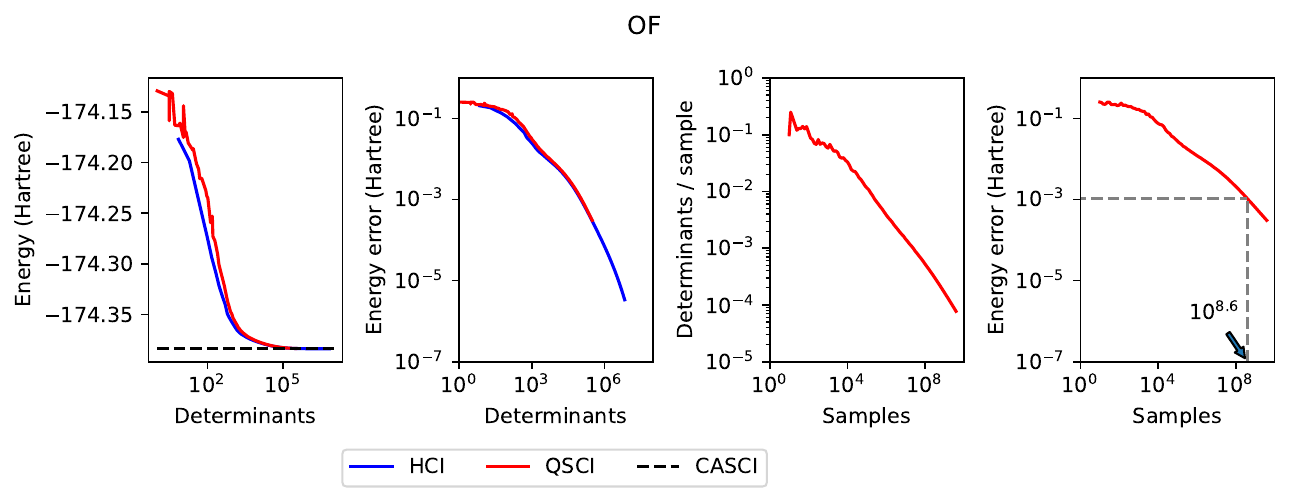}
    \caption{Ground state energy of the OF radical (doublet) in an (11e,20o) active space.
    The two left panels show the energy (linear scale) and energy error relative to the CASCI reference (log scale).
    The panels to the right show the determinants per sample and the energy error (log scale) as a function of the number of samples. 
    The small arrow on the rightmost panel indicates the number of samples required to reach milli-Hartree precision. }
\end{figure}

\begin{figure}
\centering
\includegraphics[width=16.0cm]{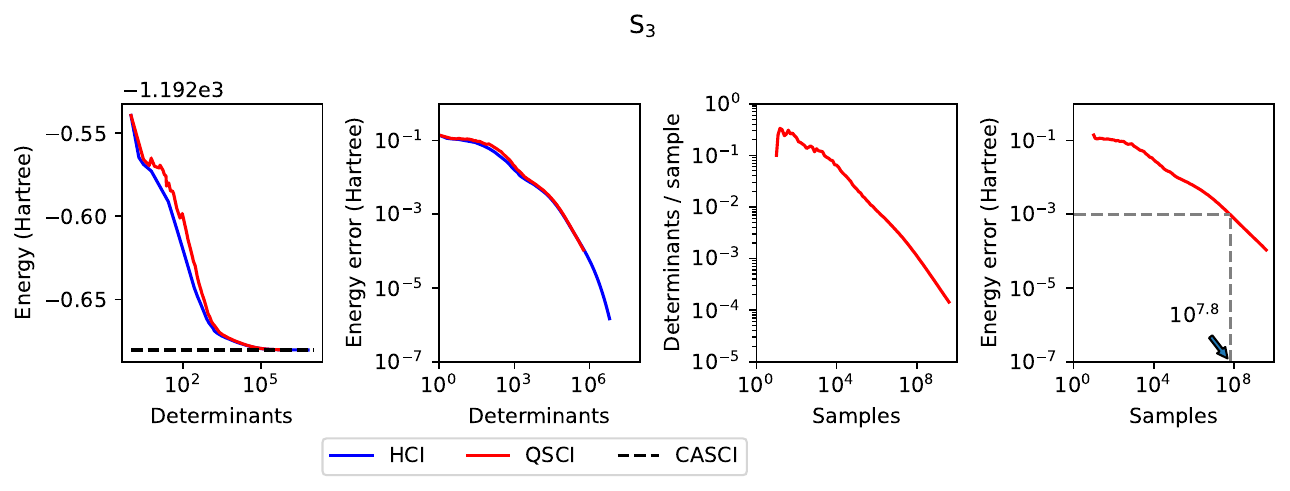}
    \caption{Ground state energy of the S$_3$ molecule in an (18e,16o) active space.
    The two left panels show the energy (linear scale) and energy error relative to the CASCI reference (log scale).
    The panels to the right show the determinants per sample and the energy error (log scale) as a function of the number of samples. 
    The small arrow on the rightmost panel indicates the number of samples required to reach milli-Hartree precision. }
\end{figure}

\begin{figure}
\centering
\includegraphics[width=16.0cm]{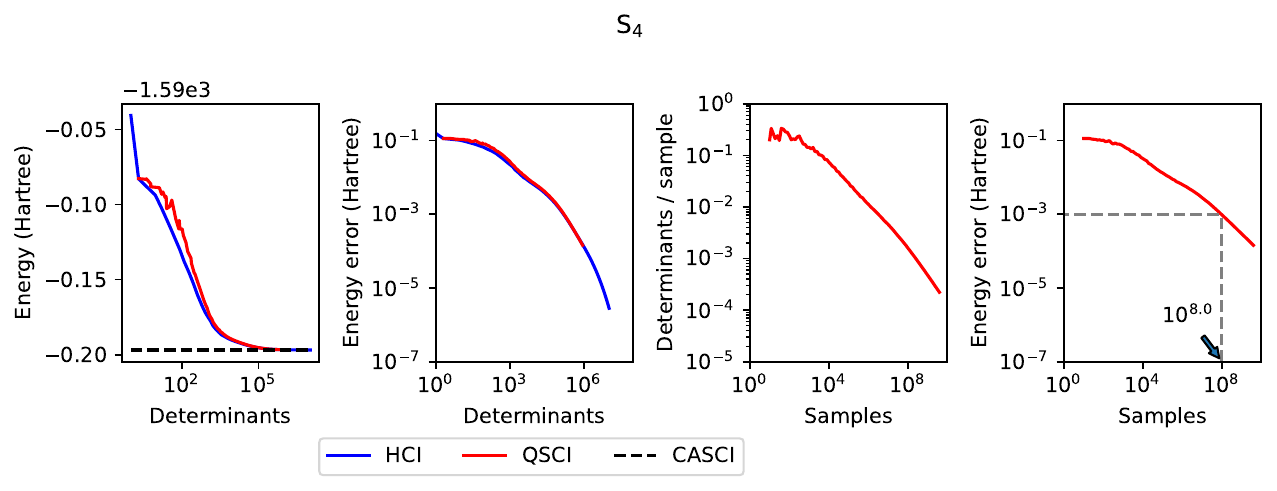}
    \caption{Ground state energy of the S$_4$ molecule in an (24e,18o) active space.
    The two left panels show the energy (linear scale) and energy error relative to the CASCI reference (log scale).
    The panels to the right show the determinants per sample and the energy error (log scale) as a function of the number of samples. 
    The small arrow on the rightmost panel indicates the number of samples required to reach milli-Hartree precision. }
\end{figure}

\newpage
\FloatBarrier
\subsection{Hand-picked systems}
\FloatBarrier

\begin{figure}
    \centering
\includegraphics[width=16.0cm]{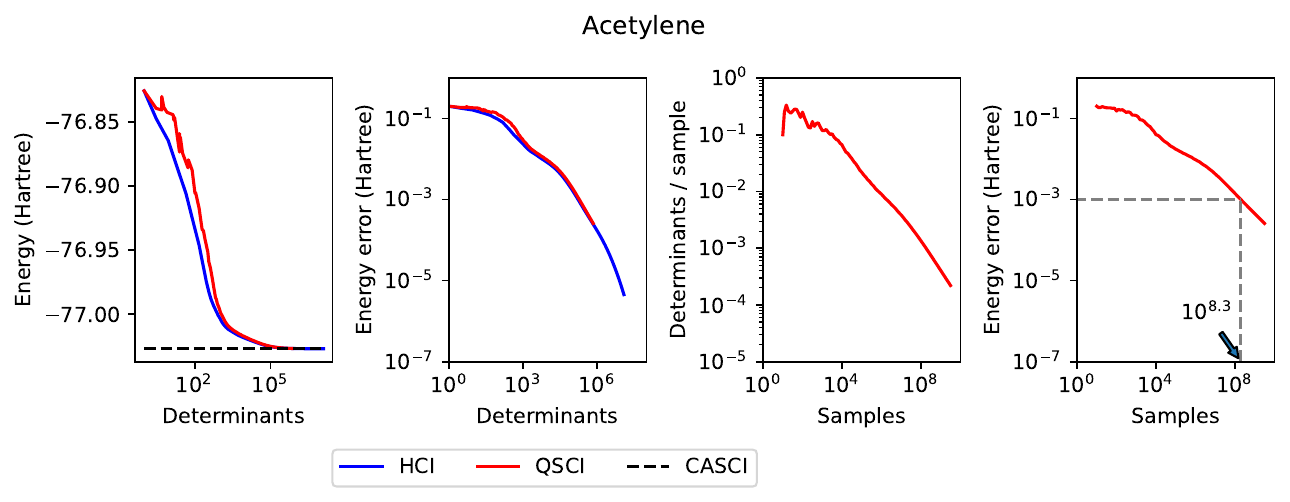}
    \caption{Ground state energy of the acetylene molecule in an (10e,22o) active space. The two left panels show the energy (linear scale) and energy error relative to the CASCI reference (log scale). The panels to the right show the determinants per sample and the energy error (log scale) as a function of the number of samples. The small arrow on the rightmost panel indicates the number of samples required to reach milli-Hartree precision. }
\end{figure}
\begin{figure}
    \centering
\includegraphics[width=16.0cm]{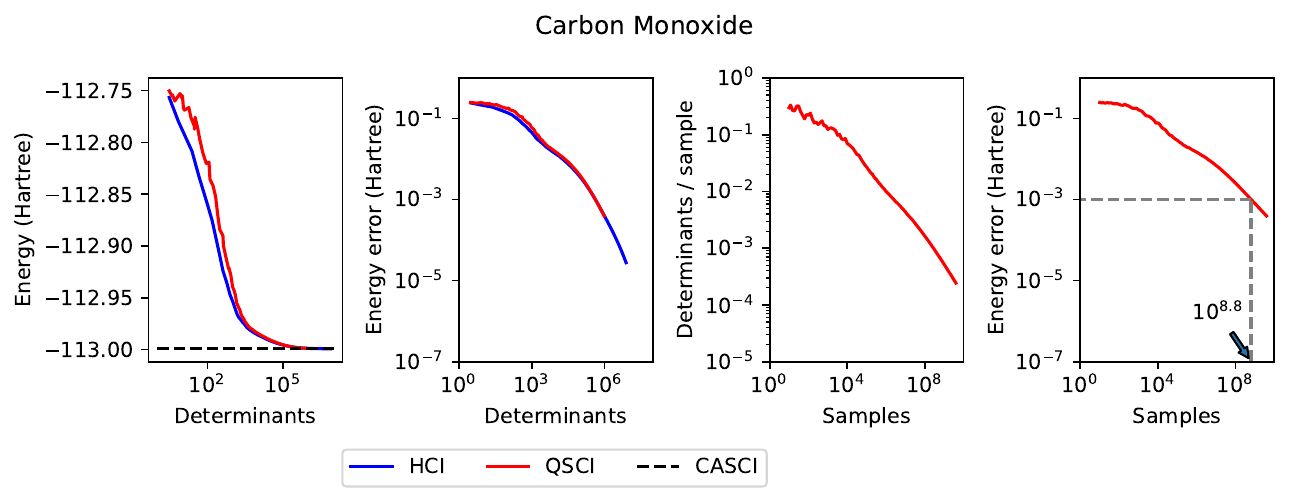}
    \caption{Ground state energy of carbon monoxide in an (10e,21o) active space. The two left panels show the energy (linear scale) and energy error relative to the CASCI reference (log scale). The panels to the right show the determinants per sample and the energy error (log scale) as a function of the number of samples. The small arrow on the rightmost panel indicates the number of samples required to reach milli-Hartree precision. }
\end{figure}
\begin{figure}
    \centering
\includegraphics[width=16.0cm]{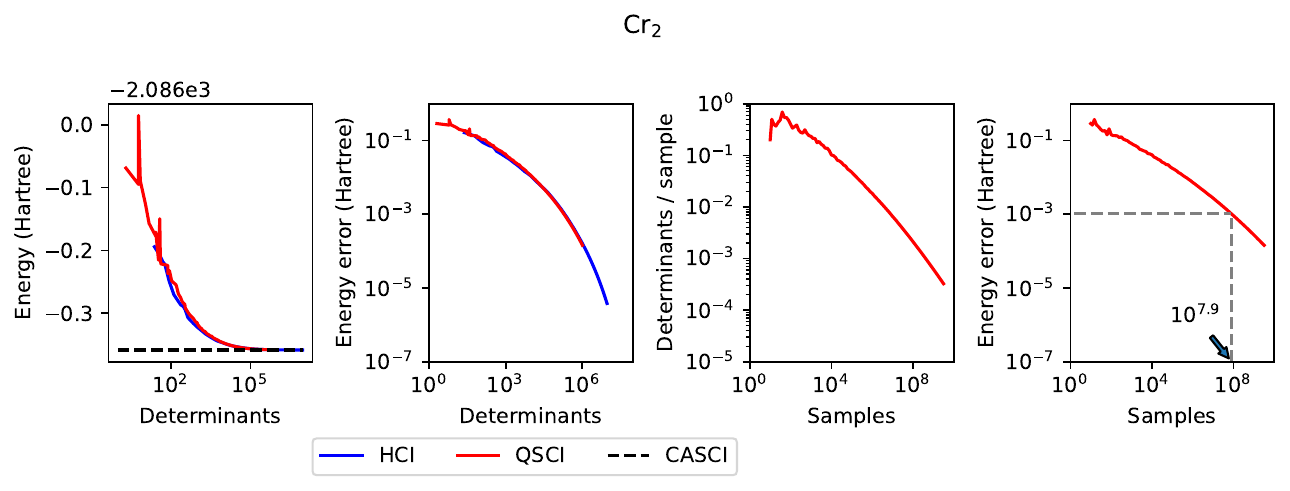}
    \caption{Ground state energy of the chromium dimer (equilibrium geometry) in an (12e,19o) active space. The two left panels show the energy (linear scale) and energy error relative to the CASCI reference (log scale). The panels to the right show the determinants per sample and the energy error (log scale) as a function of the number of samples. The small arrow on the rightmost panel indicates the number of samples required to reach milli-Hartree precision. }
\end{figure}

\begin{figure}
    \centering
\includegraphics[width=16.0cm]{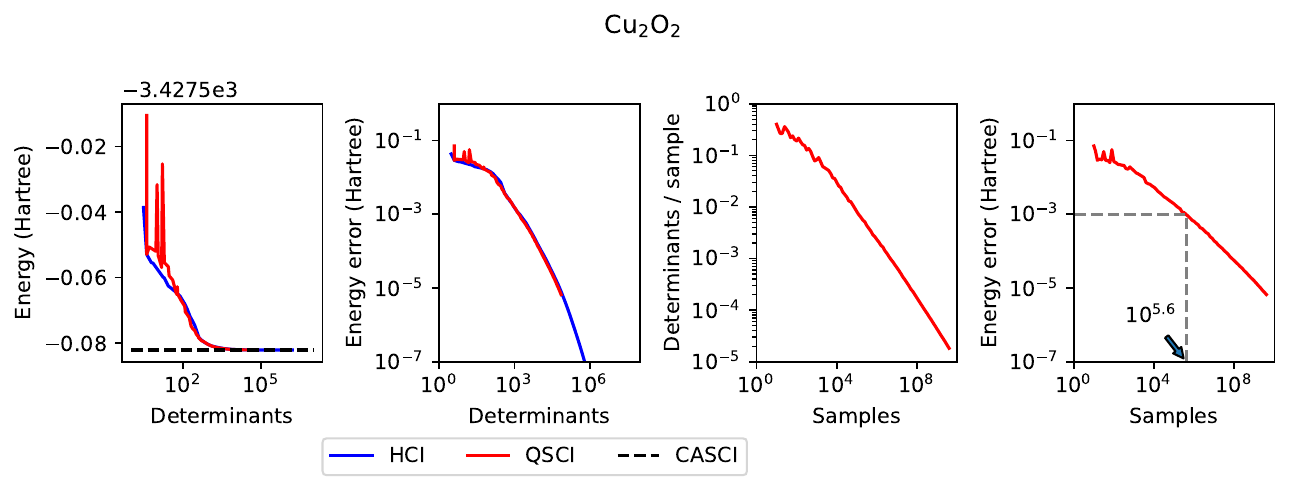}
    \caption{Ground state energy of the Cu$_2$O$_2$ molecule in an (30e,20o) active space. The two left panels show the energy (linear scale) and energy error relative to the CASCI reference (log scale). The panels to the right show the determinants per sample and the energy error (log scale) as a function of the number of samples. The small arrow on the rightmost panel indicates the number of samples required to reach milli-Hartree precision. }
\end{figure}
\begin{figure}
    \centering
\includegraphics[width=16.0cm]{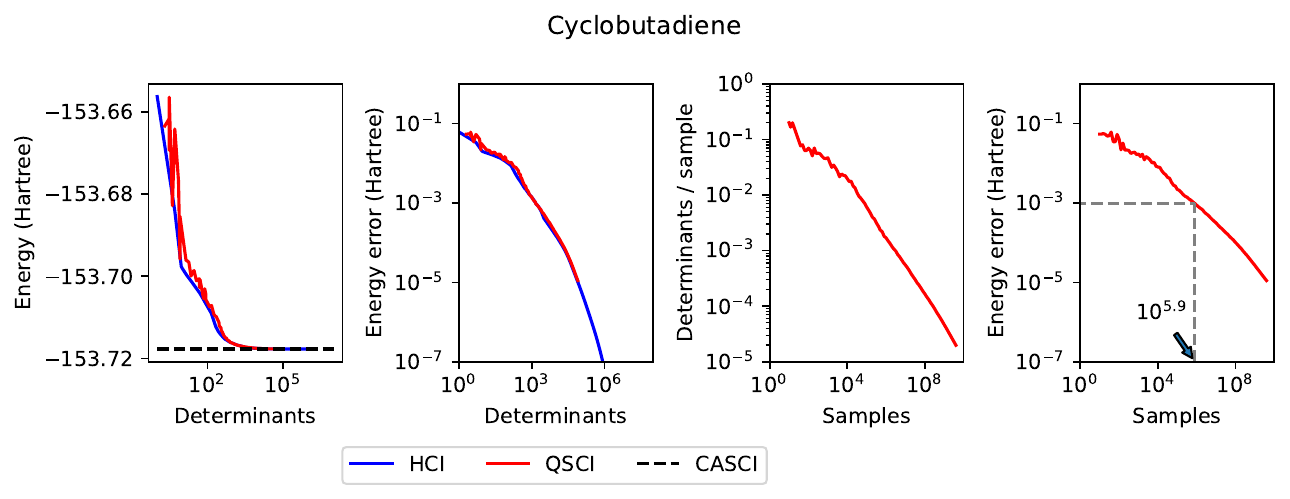}
    \caption{Ground state energy of the cyclobutadiene in an (20e,16o) active space. The two left panels show the energy (linear scale) and energy error relative to the CASCI reference (log scale). The panels to the right show the determinants per sample and the energy error (log scale) as a function of the number of samples. The small arrow on the rightmost panel indicates the number of samples required to reach milli-Hartree precision. }
\end{figure}
\begin{figure}
    \centering
\includegraphics[width=16.0cm]{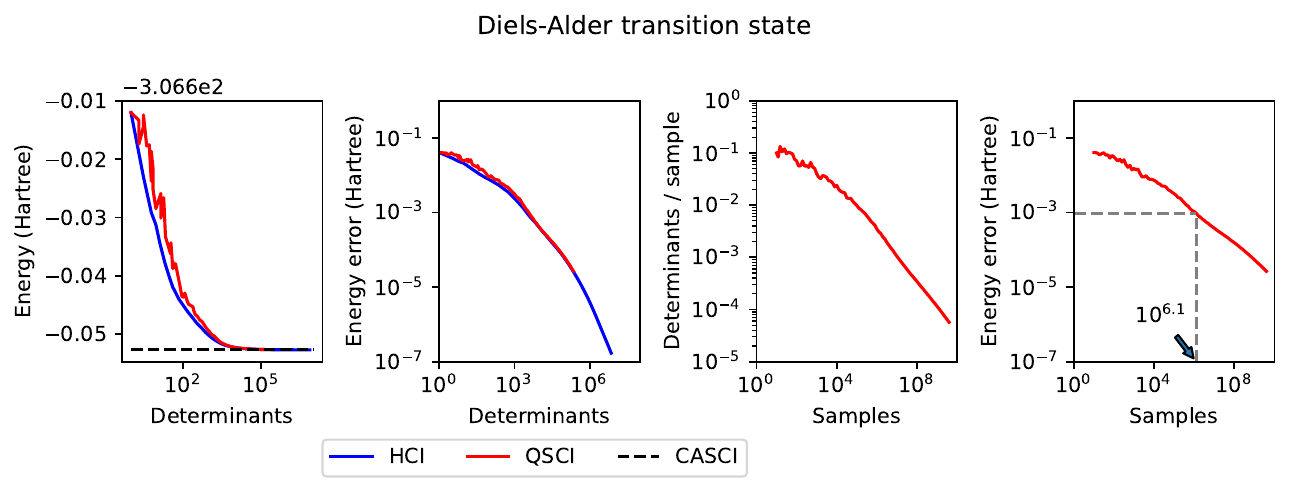}
    \caption{Ground state energy of a Diels-Alder reaction (between furan and ethene) transition state in an (16e,17o) active space. The two left panels show the energy (linear scale) and energy error relative to the CASCI reference (log scale). The panels to the right show the determinants per sample and the energy error (log scale) as a function of the number of samples. The small arrow on the rightmost panel indicates the number of samples required to reach milli-Hartree precision. }
\end{figure}
\begin{figure}
    \centering
\includegraphics[width=16.0cm]{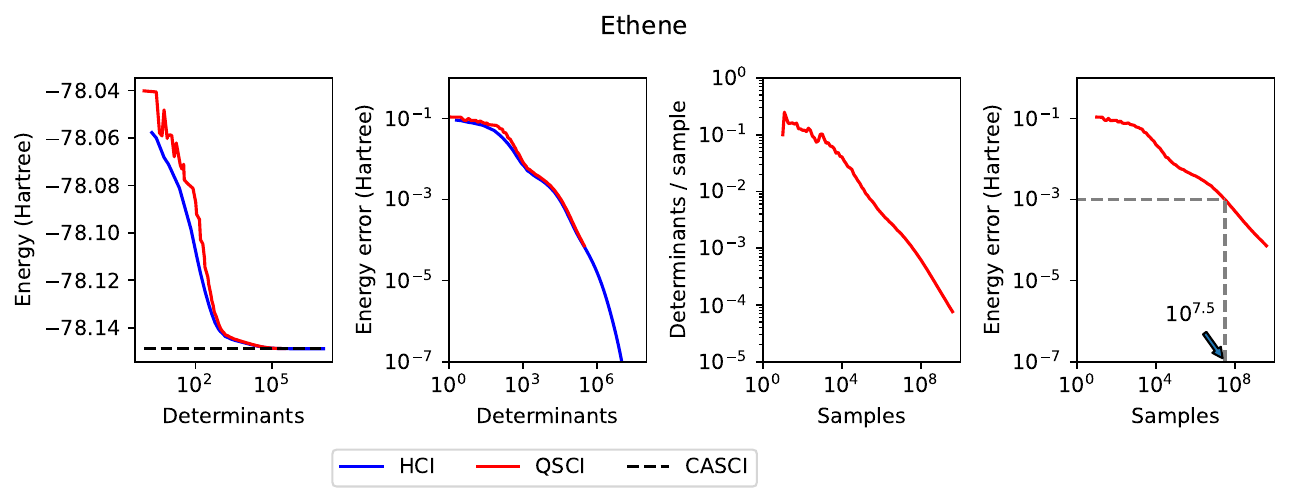}
    \caption{Ground state energy of the ethene molecule in an (12e,18o) active space. The two left panels show the energy (linear scale) and energy error relative to the CASCI reference (log scale). The panels to the right show the determinants per sample and the energy error (log scale) as a function of the number of samples. The small arrow on the rightmost panel indicates the number of samples required to reach milli-Hartree precision. }
\end{figure}
\begin{figure}
    \centering
\includegraphics[width=16.0cm]{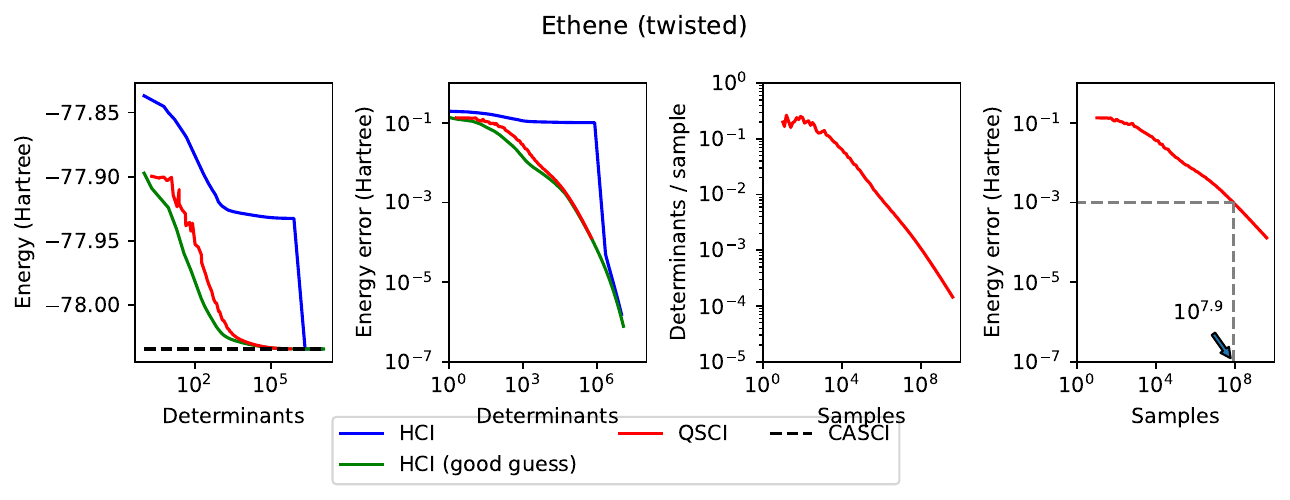}
    \caption{Ground state energy of the ethene molecule (rotated to 90 degrees along the C=C bond) in an (12e,18o) active space. The two left panels show the energy (linear scale) and energy error relative to the CASCI reference (log scale). The panels to the right show the determinants per sample and the energy error (log scale) as a function of the number of samples. The small arrow on the rightmost panel indicates the number of samples required to reach milli-Hartree precision. The "good" guess (green lines) for HCI uses a singly-excited HOMO-LUMO determinant as the initial CI expansion.}
\end{figure}
\begin{figure}
    \centering
\includegraphics[width=16.0cm]{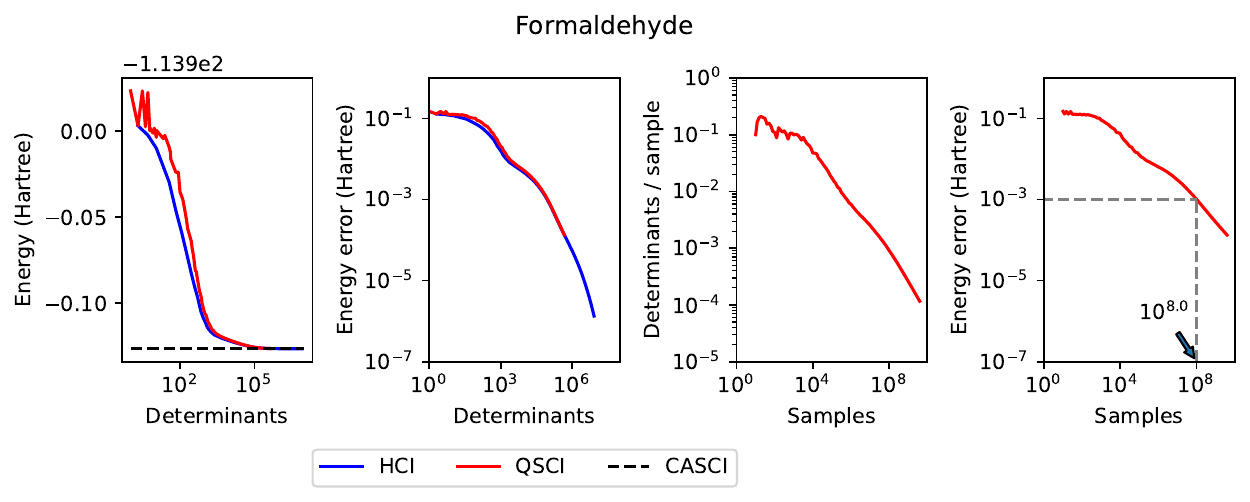}
    \caption{Ground state energy of the formaldehyde molecule in an (12e,18o) active space. The two left panels show the energy (linear scale) and energy error relative to the CASCI reference (log scale). The panels to the right show the determinants per sample and the energy error (log scale) as a function of the number of samples. The small arrow on the rightmost panel indicates the number of samples required to reach milli-Hartree precision. }
\end{figure}
\begin{figure}
    \centering
\includegraphics[width=16.0cm]{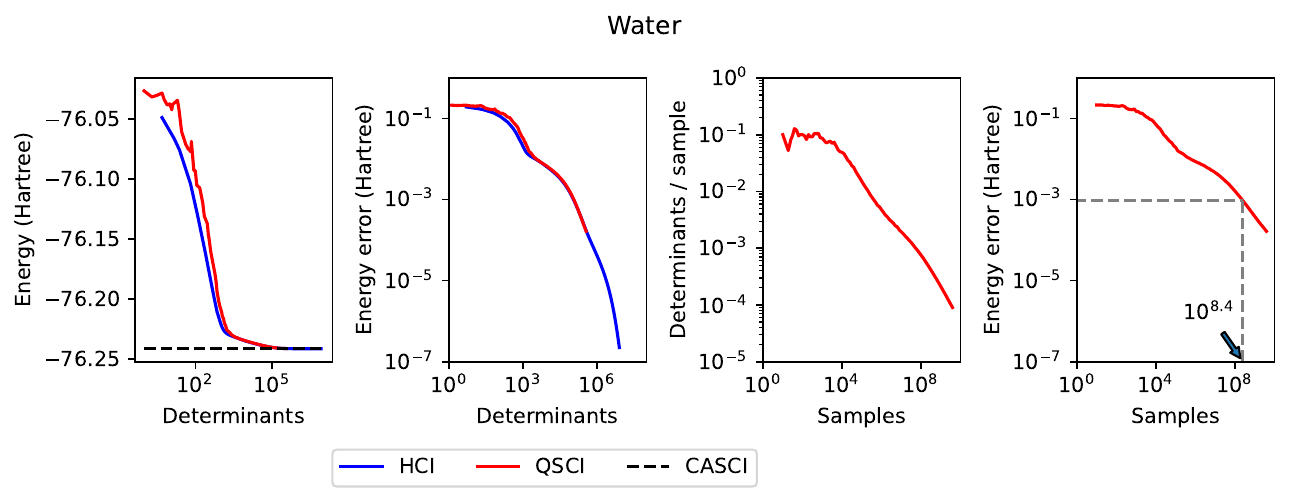}
    \caption{Ground state energy of the water molecule in an (8e,23o) active space. The two left panels show the energy (linear scale) and energy error relative to the CASCI reference (log scale). The panels to the right show the determinants per sample and the energy error (log scale) as a function of the number of samples. The small arrow on the rightmost panel indicates the number of samples required to reach milli-Hartree precision. }
\end{figure}
\begin{figure}
    \centering
\includegraphics[width=16.0cm]{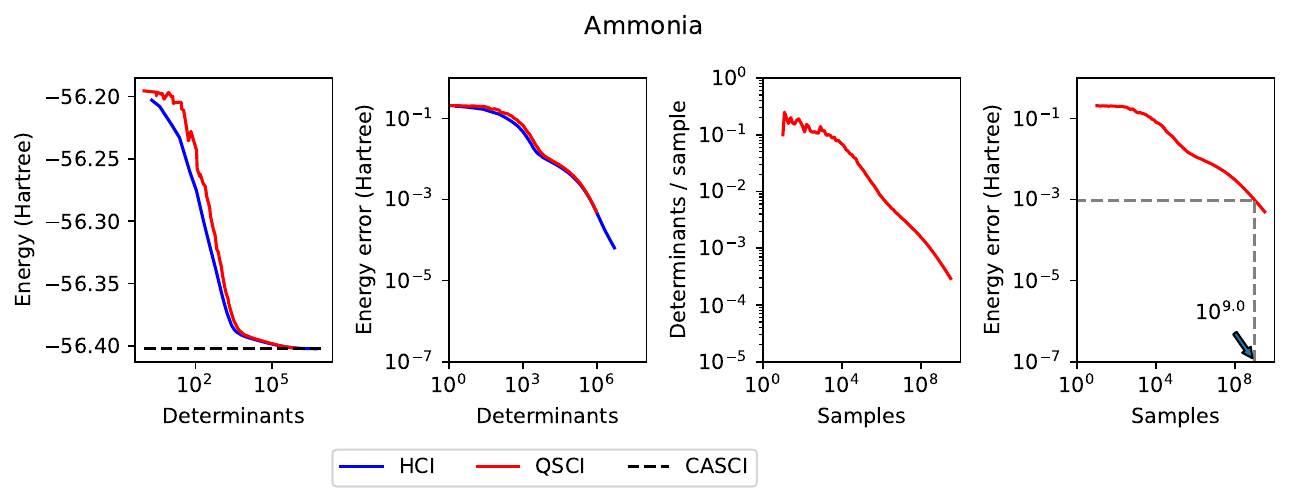}
    \caption{Ground state energy of the ammonia molecule in an (8e,28o) active space. The two left panels show the energy (linear scale) and energy error relative to the CASCI reference (log scale). The panels to the right show the determinants per sample and the energy error (log scale) as a function of the number of samples. The small arrow on the rightmost panel indicates the number of samples required to reach milli-Hartree precision. }
\end{figure}
\begin{figure}
    \centering
\includegraphics[width=16.0cm]{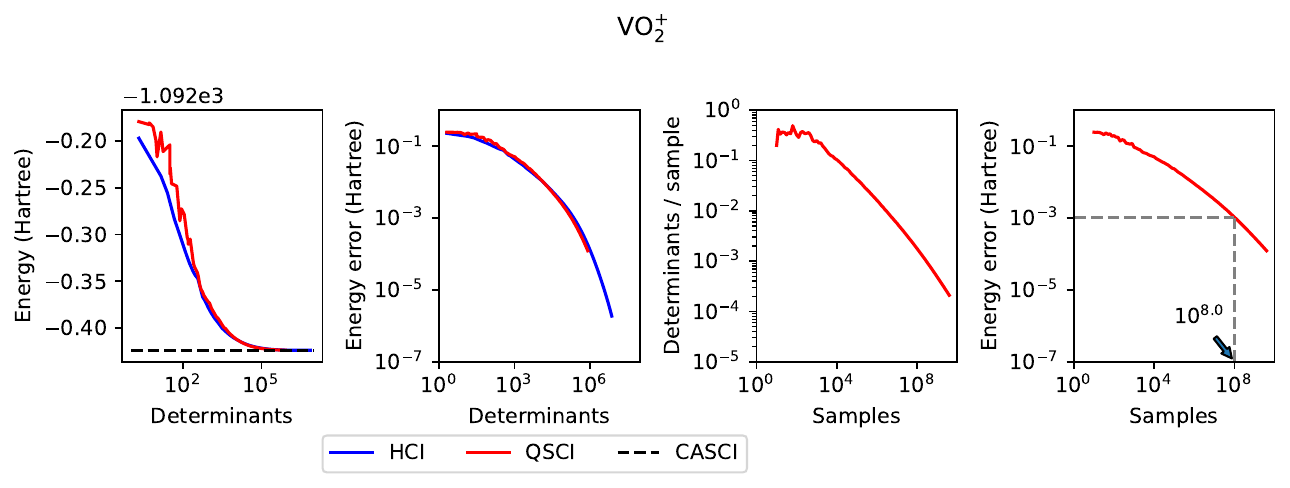}
    \caption{Ground state energy of the VO$_2^{+}$ ion in an (12e,18o) active space. The two left panels show the energy (linear scale) and energy error relative to the CASCI reference (log scale). The panels to the right show the determinants per sample and the energy error (log scale) as a function of the number of samples. The small arrow on the rightmost panel indicates the number of samples required to reach milli-Hartree precision. }
\end{figure}

\FloatBarrier
\bibliography{main}